\documentclass[ footinbib,
                twocolumn,
                showpacs,
                reprint,
                floatfix,
                superscriptaddress,
                aip,
								rsi,
                a4paper,
                raggedbottom,
								final,
                citeautoscript]{revtex4-1}

\usepackage{pdfsync}
\usepackage{subfigure}
\usepackage{graphicx}
\usepackage{amsmath, amssymb, bm}
\usepackage{dcolumn}
\usepackage{color}
\usepackage{lmodern}
\usepackage{pifont}

\newcommand{\sss}{\scriptscriptstyle}
\newcommand{\sst}{\scriptstyle}

\newcommand{\stext}[1]{\sss \text{#1} \sst}

\newcommand{\eps}{\ensuremath{\text{\large $\boldsymbol{\varepsilon}$}}}
\newcommand{\GMA}{\ensuremath{\text{\large $\boldsymbol{\gamma}$}}}

\begin{document}

\title{An integrated mid-infrared, far-infrared and terahertz optical Hall effect instrument}

\author{P.~K\"uhne}
\email{\\kuehne@huskers.unl.edu}
\homepage{(http://ellipsometry.unl.edu/)}
\affiliation{Department of Electrical Engineering and Center for Nanohybrid Functional Materials, University of Nebraska-Lincoln, Lincoln, Nebraska 68588, USA}

\author{C.~M.~Herzinger}
\affiliation{J. A. Woollam Co., Inc., 645 M Street, Suite 102, Lincoln, Nebraska 68508-2243, USA}

\author{M.~Schubert}
\affiliation{Department of Electrical Engineering and Center for Nanohybrid Functional Materials, University of Nebraska-Lincoln, Lincoln, Nebraska 68588, USA}

\author{J.A.~Woollam}
\affiliation{J. A. Woollam Co., Inc., 645 M Street, Suite 102, Lincoln, Nebraska 68508-2243, USA}

\author{T.~Hofmann}
\affiliation{Department of Electrical Engineering and Center for Nanohybrid Functional Materials, University of Nebraska-Lincoln, Lincoln, Nebraska 68588, USA}

\date{\today}

\begin{abstract}
We report on the development of the first integrated mid-infrared, far-infrared and terahertz optical Hall effect instrument, covering an ultra wide spectral range from 3~cm$^{-1}$ to 7000~cm$^{-1}$ (0.1--210~THz or 0.4--870~meV). The instrument comprises four sub-systems, where the magneto-cryostat-transfer sub-system enables the usage of the magneto-cryostat sub-system with the mid-infrared ellipsometer sub-system, and the far-infrared/terahertz ellipsometer sub-system. Both ellipsometer sub-systems can be used as variable angle-of-incidence spectroscopic ellipsometers in reflection or transmission mode, and are equipped with multiple light sources and detectors. The ellipsometer sub-systems are operated in polarizer-sample-rotating-analyzer configuration granting access to the upper left $3\times 3$ block of the normalized $4\times 4$ Mueller matrix. The closed cycle magneto-cryostat sub-system provides sample temperatures between room temperature and 1.4~K and magnetic fields up to 8~T, enabling the detection of transverse and longitudinal magnetic field-induced birefringence. We discuss theoretical background and practical realization of the integrated mid-infrared, far-infrared and terahertz optical Hall effect instrument, as well as acquisition of optical Hall effect data and the corresponding model analysis procedures. Exemplarily, epitaxial graphene grown on 6\textit{H}-SiC, a tellurium doped bulk GaAs sample and an AlGaN/GaN high electron mobility transistor structure are investigated.
The selected experimental datasets display the full spectral, magnetic field and temperature range of the instrument and demonstrate data analysis strategies. Effects from free charge carriers in two dimensional confinement and in a volume material, as well as quantum mechanical effects (inter-Landau-level transitions) are observed and discussed exemplarily.
\end{abstract}

\pacs{}

\maketitle

\section{Introduction}\label{sec:Introduction}
The optical Hall effect (OHE) is a physical phenomenon, which describes transverse and longitudinal magnetic field-induced birefringence, caused by the non-reciprocal magneto-optic response of electric charge carriers. 
The term OHE is used in analogy to the classic, electrical Hall effect\cite{HallAJM2_1879}, since the electrical Hall effect and certain cases of OHE observation can be explained by extensions of the classic Drude model for the transport of electrons in matter (metals).\cite{Drude00,ANDP:ANDP19003060312} For the OHE, Drude's classic model is extended by a magnetic field and frequency dependency, describing the electron's momentum under the influence of the Lorentz force. As a result an antisymmetric contribution is added to the dielectric tensor $\eps(\omega)$, whose sign depends on the type of the free charge carrier (electron, hole). The non-vanishing off-diagonal elements of the dielectric tensor reflect the magneto-optic birefringence, which lead to conversion of \textit{p}-polarized into \textit{s}-polarized electromagnetic waves, and vice versa. 

The OHE can be quantified in terms of the Mueller matrix, which characterizes the transformation of an electromagnetic wave's polarization state.\cite{Hecht_1987} Experimentally the Mueller matrix is measured by generalized ellipsometry (GE).\cite{AzzamBook_1984,SchubertJOSAA13_1996,Schubert96,Tiwald00,Schubert02,Schubert03a,Thompkins_2004,SchubertIRSEBook_2004} During a GE measurement different polarization states of the incident light are prepared and their change upon reflection from or transmission through a sample is determined.

An OHE instrument conducts GE measurements on samples in high, quasi-static magnetic fields, and detects the magnetic field induced changes of the Mueller matrix.\cite{Hofmannpss205_2008} Though several instruments with partial OHE instrument characteristics were described in the literature, most instruments did not fulfill all criteria for an OHE instrument. Nederpel and Martens developed in 1985 a single wavelength~(444~nm) magneto-optical ellipsometer for the visible spectral range, but the instrument provided only low magnetic fields ($B\leq 50$~mT).\cite{RSI_vis_OHE} 
In 2003 \v{C}erne \textit{et~al.} presented a magneto-polarimetry instrument ($B\leq 8$~T) for the mid-infrared spectral range~(spectral lines of CO$_2$ laser)\cite{RSI_2003_MO-polarimetry}, and in 2004 Padilla \textit{et~al.} developed a terahertz-visible (6 to 20000~cm$^{-1}$) magneto-reflectance and -transmittance instrument ($B\leq 9$~T)\cite{BasovRSI_2004}. While both instruments provide high magnetic fields, and contain polarizers and photo-elastic-modulators, these instruments were not designed to record Mueller matrix data (GE). The full $4\times 4$~Mueller matrix in the terahertz-mid-infrared spectral range (20 to 4000~cm$^{-1}$) can be measured by an instrument described in 2013 by Stanislavchuk \textit{et~al.},\cite{SirenkoRSI_2013} but here the instrument is not designed for experiments with the sample in magnetic fields.

The first full OHE instrument developed in 2006 by Hofmann \textit{et~al.} for the far-infrared (FIR) spectral range (30 to 650~cm$^{-1}$) provided magnetic fields up to $6$~T and allowed sample temperatures between $4.2$~K and room temperature.\cite{HofmannRSI77_2006} This first OHE instrument has since been successfully used to determine free charge carrier properties\cite{Hofmann02,Hofmann03,HofmannAPL82_2003,SchubertJOSAA20_2003,SchubertTSF455-456_2004,schoecheJAP_2013}, including effective mass parameters for a variety of material systems.\cite{SchubertJOSAA20_2003,HofmannPSSC3_2006,HofmannAPL88_2006,HofmannAPL90_2007,HofmannJEM_2008} Later, OHE experiments were conducted in the terahertz (THz) spectral range,\cite{HofmannRSI81_2010} but were limited to room temperature and low magnetic fields \mbox{($B\leq 1.8$~T)}.\cite{HofmannTSF519_2011,KuhneTSFxx_2011,HofmannAPL98_2011,SchocheAPL98_2011} Since the magnitude of the OHE depends on the magnetic field strength, higher magnetic fields facilitate the detection of the OHE. Furthermore, the sensitivity to the OHE is greatly enhanced by phonon mode coupling,\cite{KuehneMRS_2013,SchoecheAPL103_2013} surface guided waves\cite{HofmannAPL90_2007} and Fabry-P\'erot interferences.\cite{KuhneTSFxx_2011,HofmannAPL_2012} Since these effects appear from the THz to the mid-infrared (MIR) spectral range, depending on the structure and material of the sample, it is necessary to extend the spectral range covered by OHE instrumentation. An OHE instrument for the MIR, for example, can detect the magneto-optic response of free charge carriers enhanced by phonon modes present in the spectral range above 600 cm$^{-1}$, which applies to many substrate materials, \textit{e.g.}, SiC,\cite{Tiwald99b,KuehneMRS_2013,KuehnePRL111_2013} Al$_2$O$_3$,\cite{SchubertPRB61_2000,HofmannJEM_2008} or GaN,\cite{KasicPRB62_2000} as well as to many materials used for thin films, \textit{e.g.}, III-V nitride semiconductors Al$_{1-x}$Ga$_{x}$N,\cite{Kasic02b,SchoecheAPL103_2013} Al$_{1-x}$In$_{x}$N\cite{Kasic03} or In$_{1-x}$Ga$_{x}$N\cite{Kasic03,HofmannJEM_2008}. In addition, inter-Landau-level transitions can be studied in the MIR spectral range\cite{PhysRevLett.20.1292,PhysRevB.86.155409,SadowskiACP893_2007,SadowskiSSC143_2007} with a MIR OHE instrument.\cite{KuehnePRL111_2013} The extension to the THz spectral range enables the detection of the OHE in samples with low carrier concentrations.\cite{HofmannRSI77_2006,KuhneTSFxx_2011} Furthermore, the strongest magneto-optic response can be observed at the cyclotron resonance frequency, which typically lies in the microwave/THz spectral range for moderate magnetic fields (few Tesla) and effective mass values comparable to the free electron mass.

In this article, we present an OHE instrument, covering an ultra wide spectral range from 3~cm$^{-1}$ to 7000~cm$^{-1}$ (0.1--210~THz or 0.4--870~meV), which combines MIR, FIR and THz magneto-optic generalized ellipsometry in a single instrument. This integrated MIR, FIR and THz OHE instrument incorporates a commercially available, closed cycle refrigerated, superconducting 8~Tesla magnet-cryostat sub-system, with four optical ports, providing sample temperatures between $T=1.4$~K and room temperature. The ellipsometer sub-systems were built in-house and operate in the rotating-analyzer configuration, capable of determining the normalized upper $3 \times 3$ block of the sample Mueller matrix. 



The operation of the integrated MIR, FIR and THz OHE instrument is demonstrated by three sample systems. Combined experimental data from the MIR, FIR and THz spectral range of a single epitaxial graphene sample, grown on a 6\textit{H}-SiC substrate by thermal decomposition, are shown. The MIR OHE data of the same  epitaxial graphene sample is used to demonstrate the operation of the MIR ellipsometer sub-system of the integrated MIR, FIR and THz OHE instrument,\cite{KuehnePRL111_2013} over the full available magnetic field range of the instrument. The magneto-optic response of free charge carriers and quantum mechanical inter-Landau-level transitions is observed, and their polarization selection rules obtained therefrom are briefly discussed. A Te-doped, n-type GaAs substrate serves as a model system for the FIR spectral range of the FIR/THz ellipsometer sub-system. The OHE signal originating from conduction band electrons in a bulk material is discussed, and the concentration, mobility, and effective mass of the conduction band electrons is determined. Finally, OHE data from an AlGaN/GaN high electron mobility transistor structure (HEMT) from the THz spectral range of the FIR/THz ellipsometer sub-system is presented and analyzed.\cite{HofmannAPL_2012} The data was recorded at different temperatures between $T=1.5$~K and room temperature, representing the full sample temperature range of the instrument.

The manuscript is organized as follows, in section~\ref{sec:Theory} dielectric and magneto-optic dielectric tensors are introduced, a brief theoretical overview on Mueller matrices and GE data-acquisition is given, and general GE data analysis procedures are introduced. Section~\ref{sec:Experimental setup} gives a detailed description of the experimental setup, while in section~\ref{sec:ACQ-ANAL} data acquisition and data analysis procedures for OHE data are discussed. Examples of experimental results, demonstrating the operation of the integrated MIR, FIR and THz OHE instrument, are presented and discussed in section~\ref{sec:samplesystems}, which is followed by a short summary in section~\ref{sec:Summary}.

\section{Theory}\label{sec:Theory}
The evaluation of physical relevant parameters from the OHE requires the experimental observation and quantification of the OHE, and a physical model to analyze OHE data. Experimentally, the OHE is quantified in terms of the Mueller matrix $\mathbf{M}_{\text{\tiny{OHE}}}$\cite{Fujiwara_2007,goldstein2011polarized} by employing generalized ellipsometry (GE). The physical model which is used to analyze the observed transverse and longitudinal magneto-optic birefringence of the OHE is based on the magneto-optic dielectric tensor $\eps_{\text{\tiny{OHE}}}(\mathbf{B})$, which is a function of the slowly varying external magnetic field $\mathbf{B}$. If, among other parameters, the magneto-optic dielectric tensor of a sample is known, experimental Mueller matrices $\mathbf{M}_{\text{\tiny{OHE}}}$ can be modeled from $\eps_{\text{\tiny{OHE}}}(\mathbf{B})$ using the relation
\begin{equation}
	\mathbf{M}_{\text{\tiny{OHE}}}(\eps_{\text{\tiny{OHE}}}(\mathbf{B}))\;.
	\label{eqn:MM-DF}
\end{equation}
This relation is in general not invertible analytically, but can be used to determine the magneto-optic dielectric tensor from experimental Mueller matrix data through non-linear model regression analysis\cite{Rossler99}. Dielectric tensors, Mueller matrix calculus, generalized ellipsometry including data acquisition, as well as data analysis will be addressed in this section. 

\subsection{Magneto-optical dielectric tensors}\label{sec:DFs}
The optical response of a sample is here described by the dielectric tensor $\eps$. If the dielectric tensor of the sample without a magnetic field is given by $\eps_{_{\hspace{-1pt}\mathbf{B}=0}}$ and the change of the dielectric tensor induced by a magnetic field $\mathbf{B}$ by $\eps_{_{\hspace{-1pt}\mathbf{B}}}$, the magneto-optic dielectric tensor describing the OHE, can be expressed as
\begin{equation}
		\eps_{\text{\tiny{OHE}}}(\mathbf{B})
		=
		\eps_{_{\hspace{-1pt}\mathbf{B}=0}}
		+\;
		\eps_{_{\hspace{-1pt}\mathbf{B}}}\;.
\end{equation}
The magneto-optic response of the sample described by $\eps_{_{\hspace{-1pt}\mathbf{B}}}$ usually originates from bound and unbound charge carriers subjected to the magnetic field and is caused by the action of the Lorentz force. The magneto-optic response is anisotropic, and non-reciprocal in time.\cite{SchubertIRSEBook_2004,Hofmann05Thesis} Thus, the corresponding magneto-optic contributions $\chi_{\mathrm{+}}$ and $\chi_{\mathrm{-}}$ to the permittivity tensor $\bm{\chi}=\eps-\mathbf{I}$, where $\mathbf{I}$ is the $3\times 3$ identity matrix, originate from the interaction of right- and left-handed circularly polarized light with the sample, respectively.\cite{SchubertJOSAA20_2003, Hofmannpss205_2008} Without loss of generality, if the magnetic field $\mathbf{B}$ is pointing in the \textit{z}-direction, the polarization vector \mbox{$\mathbf{P}=\varepsilon_0\bm{\chi}\mathbf{E}$} can be described by arranging the electric fields in their circularly polarized eigensystem \mbox{$\mathbf{E}_e = (E_x + \mathrm{i}E_y, E_x - \mathrm{i}E_y, E_z) = (E_{+}, E_{-}, E_z)$} by $\mathbf{P}_e = \varepsilon_0\bm{\chi}_e \mathbf{E}_e = \varepsilon_0(\chi_{\mathrm{+}}E_{+}, \chi_{\mathrm{-}}E_{-}, 0)$, where $\text{i}=\sqrt{-1}$ is the imaginary unit.\cite{SchubertADP15_2006,HofmannRSI77_2006} Transforming $\mathbf{P}_e$ back into the laboratory system the change of the dielectric tensor induced by the magnetic field takes the form:\cite{SchubertADP15_2006, HofmannRSI77_2006} \footnote{The identity matrix $\mathbf{I}$ is not part of the magneto-optic dielectric tensor $\eps_{_{\hspace{-1pt}\mathbf{B}}}$ since it is already incorporated in the non-magnetic part of the dielectric tensor $\eps_{_{\hspace{-1pt}\mathbf{B}=0}}$.}
\begin{equation} 
	\label{eq:LLTeps}
	\eps_{_{\hspace{-1pt}\mathbf{B}}}
	=
	\frac{1}{2}
	\begin{pmatrix}
	\hspace{11pt}(\chi_{\textbf{\tiny{+}}}+\chi_{\textbf{\tiny{--}}}) & \text{i}(\chi_{\textbf{\tiny{+}}}-\chi_{\textbf{\tiny{--}}}) & 0 \\
	   -\text{i}(\chi_{\textbf{\tiny{+}}}-\chi_{\textbf{\tiny{--}}}) & \hspace{3pt}(\chi_{\textbf{\tiny{+}}}+\chi_{\textbf{\tiny{--}}}) & 0 \\
			0																						& 0																						& 0 \\
	\end{pmatrix}\;.
\end{equation}
Note, under field inversion $\mathbf{B}\rightarrow -\mathbf{B}$, the polarizabilities for left- and right-handed circularly polarized light interchange. $\eps_{_{\hspace{-1pt}\mathbf{B}}}$ is only diagonal if $\chi_{\mathrm{+}}=\chi_{\mathrm{-}}$, and otherwise is non-diagonal with anti-symmetric off diagonal elements.

\subsubsection{Classic dielectric tensors (Lorentz-Drude model)}
Charged carriers, subject to a slowly varying magnetic field obey the classical Newtonian equation of motion (Lorentz-Drude model)\cite{Yu99}
	\begin{equation}
	\mathbf{m}\mathbf{\ddot{x}}
	+
	\mathbf{m}\GMA\mathbf{\dot{x}}
	+
	\mathbf{m} \omega_0^2 \mathbf{x}
	=
	q \mathbf{E} + q (\mathbf{\dot{x}} \times \mathbf{B})\;,
	\label{eqn:fcc-newton}
	\end{equation}
where $\mathbf{m}$, $q$, $\boldsymbol{\mu}=q\mathbf{m}^{-1}\GMA^{-1}$, $\mathbf{x}$ and $\omega_0$ represent the effective mass tensor, the electric charge, the mobility tensor, the spatial coordinate of the charged carrier and the eigenfrequency of the undamped system without external excitation and magnetic field, respectively. For a time harmonic electromagnetic plane wave with an electric field $\mathbf{E}\rightarrow\mathbf{E}\exp(\text{i}\omega t)$ with angular frequency $\omega$, the time derivative of the spatial coordinate of the charge carrier is $\mathbf{\dot{x}}=\mathbf{v}\exp(\text{i}\omega t)$, where $\mathbf{v}$ is the velocity of the charge carrier. With the current density $\mathbf{j}=n q \mathbf{v}$ Eq.~(\ref{eqn:fcc-newton}) reads
	\begin{equation}
		\mathbf{E}
		=
		\frac{1}{nq}
		\left[
			\text{i}
			\frac{\mathbf{m}}{q\omega}
			\left(
				\omega_0^2\mathbf{I}-\omega^2\mathbf{I}-\text{i}\omega\GMA
			\right)
			\mathbf{j}
			+
			(\mathbf{B} \times \mathbf{j})
		\right]\;,
	\end{equation}
where $n$ is the charge carrier density. With the Levi-Cevita-Symbol $\epsilon_{ijk}$\footnote{In the following equation the Einstein notation is used, and the covariance and contravariance is ignored since all coordinate systems are Cartesian (The summation is only executed over pairs of lower indices).}, the conductivity tensor $\mathbf{\sigma}$, the dielectric constant $\varepsilon_0$, and using \mbox{$\mathbf{E}= \boldsymbol{\sigma}^{-1} \mathbf{j}$} and \mbox{$\eps=\frac{1}{\text{i} \varepsilon_0 \omega}\boldsymbol{\sigma}$} the dielectric tensor for charge carriers subject to the external magnetic field $\mathbf{B}$ can be expressed as
\begin{equation}\label{eqn:fulldrude}
	\varepsilon_{ik}
	=
	\frac{nq^2}{\varepsilon_0}
		\left[
			m_{ik} (\omega_0^2-\omega^2-\text{i}\omega\gamma_{ik}) 
			-
			\text{i}\omega\epsilon_{ijk} q B_j
		\right]^{-1}\;.
\end{equation}

\paragraph{Polar lattice vibrations (Lorentz oscillator)}\label{sec:PhononDF}
{\quad}\newline For isotopic effective mass tensors the cyclotron frequency $\omega_\text{c}=\frac{q |B|}{m}$ can be defined. For the mass of the vibrating atoms of polar lattice vibrations, the cyclotron frequency is several orders of magnitude smaller than for effective electron masses, and can be neglected for the magnetic fields and spectral ranges discussed in this paper. Therefore, the dielectric tensor of polar lattice vibrations $\eps^{\text{\tiny{L}}}$ can be approximated using Eq.~(\ref{eqn:fulldrude}) with $\mathbf{B}=0$. When assuming isotropic effective mass and mobility tensors, the result is a simple harmonic oscillator function with Lorentzian-type broadening.\cite{Pidgeon80,Kittel86,Yu99} For materials with orthorhombic symmetry and multiple optical excitable lattice vibrations, the dielectric tensor can be diagonalized to
\begin{equation}
	\eps^{\text{\tiny{L}}}
	=
	\begin{pmatrix}
		\varepsilon_{x}^{\text{\tiny{L}}} & 0	&	0 \\
		0 & \varepsilon_{y}^{\text{\tiny{L}}}	&	0 \\
		0 & 0	&	\varepsilon_{z}^{\text{\tiny{L}}} 
	\end{pmatrix}\;,
	\label{Substrate}
\end{equation}
where $\varepsilon_{\text{\tiny{\textit{k}}}}^{\text{\tiny{L}}}$ ($k=\{x,y,z\}$) is given by\cite{Barker64}
\begin{equation}
	\varepsilon_{\text{\tiny{\textit{k}}}}^{\text{\tiny{L}}}=\varepsilon_{\infty,\text{\tiny{\textit{k}}}}\prod^{l}_{j=1}\frac{\omega^2+\text{i}\omega\gamma_{\stext{LO},\text{\tiny{\textit{k,j}}}}-\omega^2_{\stext{LO},\text{\tiny{\textit{k,j}}}}}{\omega^2+\text{i}\omega\gamma_{\stext{TO},\text{\tiny{\textit{k,j}}}}-\omega^2_{\stext{TO},\text{\tiny{\textit{k,j}}}}}\;,
	\label{eqn:phonon2}
\end{equation}
where $\omega_{\text{LO,\text{\tiny{\textit{k,j}}}}}$, $\gamma_{\text{LO,\text{\tiny{\textit{k,j}}}}}$, $\omega_{\text{TO,\text{\tiny{\textit{k,j}}}}}$, and $\gamma_{\text{TO,\text{\tiny{\textit{k,j}}}}}$ denote the \mbox{$k=\{x,y,z\}$} component of the frequency and the broadening values of the $j^{\text{th}}$ longitudinal optical (LO) and transverse optical (TO) phonon modes, respectively, while the index $j$ runs over $l$ modes. Further details can be found in Refs.~\onlinecite{Barker64,Berreman68,Gervais74, HofmannAPL88_2006, HofmannPRB66_2002}, and a detailed discussion of the requirements to the broadening parameters, such as Im$\left\{\varepsilon_{\text{\tiny{\textit{k}}}}^{\text{\tiny{L}}}\right\}\geq 0$, in Ref.~\onlinecite{KasicPRB62_2000}.

\paragraph{Free charged carriers (extended Drude model)}\label{sec:DrudeDF}
For free charged carriers no restoring force is present and the eigenfrequency of the system is $\omega_0=0$. For isotropic effective mass and conductivity tensors, and magnetic fields aligned along the \textit{z}-axis Eq.~(\ref{eqn:fulldrude}) can be written in the form
$
	\eps^{\text{\tiny{D}}}_{\text{\tiny{OHE}}}(\mathbf{B})
	=
	\eps_{_{\hspace{-1pt}\mathbf{B}=0}}^{\text{\tiny{D}}}
	+
	\eps_{_{\hspace{-1pt}\mathbf{B}}}^{\text{\tiny{D}}}
$
, with the Drude dielectric tensor for $B=0$
\begin{equation}
	\eps_{_{\hspace{-1pt}\mathbf{B}=0}}^{\text{\tiny{D}}}
	=
	-\frac{\omega_{\text{p}}^2}{\omega (\omega+\text{i}\gamma)}
	\mathbf{I}
	=
	\varepsilon^{\text{\tiny{D}}}
	\mathbf{I}\;,
\end{equation}
where $\omega_{\text{p}}=\sqrt{\frac{n q^2}{m\varepsilon_0}}$ is the plasma frequency, and $\varepsilon^{\text{\tiny{D}}}$ is the isotropic Drude dielectric function. The magneto-optic contribution to the dielectric tensor $\eps_{_{\hspace{-1pt}\mathbf{B}}}^{\text{\tiny{D}}}$ for isotropic effective masses and conductivities can be expressed, using Eq.~(\ref{eq:LLTeps}), through polarizability functions for right- and left-handed circularly polarized light
\begin{equation}
	\chi_{\mathrm{\pm}}
	=
	-\frac{\varepsilon^{\text{\tiny{D}}}}{1\mp \frac{\omega+\text{i}\gamma}{\omega_{\text{c}}}}\;,
\end{equation}
where $\omega_{\text{c}}=\frac{q |B|}{m}$ is the isotropic cyclotron frequency.

\subsubsection{Non-classic dielectric tensors (Inter-Landau-level transitions)}\label{sec:LandauDF}
The dielectric tensor $\eps_{_{\hspace{-1pt}\mathbf{B}}}^{\text{\tiny{LL}}}$ describing a series of inter-Landau-level transitions can be approximated by a sum of Lorentz oscillators. 
The quantities $\chi_{\mathrm{\pm}}$ in Eq.~(\ref{eq:LLTeps}) are then expressed by
	\begin{equation}
		\chi_{\pm} = e^{\pm\text{i}\phi}\sum_k \frac{A_{k}}{\omega^2-\omega^2_{0,k}- \text{i} \gamma_{k} \omega}\;,
		\label{eqn:landau}
	\end{equation}
where $A_{k}$, $\omega_{0,k}$, and $\gamma_{k}$ are amplitude, transition energy, and broadening parameter of the $k^{\text{th}}$ inter-Landau-level transition, respectively, which in general depend on the magnetic field. The phase factor $\phi$ was introduced empirically here to describe the experimentally observed line shapes of all Mueller matrix elements. For inter-Landau-level transitions in graphite or bi-layer graphene we find $\phi=\pi/4$, and for inter-Landau-level transitions in single layer graphene $\phi=0$.

Note that for $\phi=0$, the polarizabilities for left and right handed circularly polarized light are equal \mbox{$(\chi_{\textbf{\tiny{+}}}=\chi_{\textbf{\tiny{--}}})$}, and $\eps_{_{\hspace{-1pt}\mathbf{B}}}^{\text{\tiny{LL}}}$ is diagonal.

\subsection{Mueller matrix calculus, GE and data acquisition}\label{sec:OHE}

\subsubsection{Stokes vector/Mueller matrix calculus}\label{sec:MM-calculus}
The real-valued Stokes vector $\mathbf{S}$ has four components\footnote{Four independent components are needed to quantize all aspects of polarized light: total intensity, degree of polarization, ellipticity and orientation of the polarization ellipse.}, carries the dimension of an intensity, and can quantify any polarization state of plane electromagnetic waves. If expressed in terms of the \textit{p}- and \textit{s}-coordinate system\footnote{The letters \textit{p} and \textit{s} stand for ``\textit{parallel}'' and ``\textit{senkrecht}'' (German for parallel and perpendicular, respectively), and refer to the  directions with respect to the plane of incidence.}, its individual components can be defined by $S_1=I_p+I_s$, $S_2=I_p-I_s$, $S_3=I_{45}-I_{-45}$, and $S_4=I_{\sigma+}-I_{\mathit{\sigma}-}$, with $I_p$, $I_s$, $I_{45}$, $I_{-45}$, $I_{\sigma+}$, and $I_{\sigma-}$ being the intensities for the \textit{p}-, \textit{s}-, +45$^\circ$, -45$^\circ$, right- and left-handed circularly polarized light components, respectively.\cite{AzzamBook_1984,Roseler90}

The real-valued $4\times 4$ Mueller matrix $\mathbf{M}$ describes the change of electromagnetic plane wave properties (intensity, polarization state), expressed by a Stokes vector $\mathbf{S}$, upon change of the coordinate system or the interaction with a sample, optical element, or any other matter\cite{Fujiwara_2007,AzzamBook_1984}
\begin{equation}
	S^{(\text{out})}_j=\sum^{3}_{i=1}M_{ij}S^{(\text{in})}_i,\;\;(j=1\ldots 4)\;,
	\label{eqn:multi_MM}
\end{equation}
where $\mathbf{S}^{(\text{out})}$ and $\mathbf{S}^{(\text{in})}$ denote the Stokes vectors of the electromagnetic plane wave before and after the change of the coordinate system, or an interaction with a sample, respectively. Note that all Mueller matrix elements of the GE data discussed in this paper, are normalized by the element $M_{11}$, therefore $|M_{ij}|\leq 1$ and $M_{11}\overset{!}{\equiv}1$.


\subsubsection{Mueller matrix and OHE data}\label{sec:MM-OHE-data}
The Mueller matrix can be decomposed in 4 sub-matrices, where the matrix elements of the two off-diagonal-blocks
$
\text{
	\scriptsize
	$
		\begin{bmatrix}
			M_{13}&\hspace{-3pt}M_{14}\\
			M_{23}&\hspace{-3pt}M_{24}\\
		\end{bmatrix}
	$
	\normalsize
	and
	\scriptsize
	$
		\begin{bmatrix}
			M_{31}&\hspace{-3pt}M_{32}\\
			M_{41}&\hspace{-3pt}M_{42}\\
		\end{bmatrix}
	$
}
$
only deviate from zero if \textit{p}-\textit{s}-polarization mode conversion appears, while the matrix elements in the two on-diagonal-blocks 
$
\text{
	\scriptsize
	$
		\begin{bmatrix}
			M_{11}&\hspace{-3pt}M_{12}\\
			M_{21}&\hspace{-3pt}M_{22}\\
		\end{bmatrix}
	$
	\normalsize
	and
	\scriptsize
	$
		\begin{bmatrix}
			M_{33}&\hspace{-3pt}M_{34}\\
			M_{43}&\hspace{-3pt}M_{44}\\
		\end{bmatrix}
	$
}
$
mainly contain information about \textit{p}-\textit{s}-polarization mode conserving processes. \textit{p}-\textit{s}-polarization mode conversion is defined as the transfer of energy from the \textit{p}-polarized channel of an electromagnetic plane wave to the \textit{s}-polarized channel, or vice versa. Polarization mode conversion can appear when the \textit{p}-\textit{s}-coordinate system is different for $\mathbf{S}^{(\text{in})}$ and $\mathbf{S}^{(\text{out})}$\footnote{For example fully \textit{p}-polarized light becomes fully \textit{s}-polarized light after a 90$^{\circ}$ rotation of the coordinate system around the beam path.}, or when a sample shows birefringence, for example. In particular, polarization mode conversion appears if the dielectric tensor of a sample possesses non-vanishing off-diagonal elements. Therefore, in Mueller matrix data from optically isotropic samples, ideally all off-diagonal-block elements vanish, while, for example, magneto-optic birefringence can cause non-zero off-diagonal-block elements in the Mueller matrix.

Here, we define OHE data as Mueller matrix data from an OHE experiment [Eq.~(\ref{eqn:MM-DF})] with magnetic field $\pm \mathbf{B}$
\begin{equation}\label{OHE-MM-approx}
	\mathbf{M}^\pm_{\text{\tiny{OHE}}}
	=
	\mathbf{M}(\eps_{_{\hspace{-1pt} \mathbf{B}=0}}\hspace{-4pt}+\hspace{-2pt}\eps_{_{\hspace{-2pt}\pm \mathbf{B}}})\;.
\end{equation}
Furthermore, we define the derived OHE datasets $\delta\mathbf{M}^{\pm}$ as difference data between the Mueller matrix datasets, measured at the magnetic field $\pm \mathbf{B}$ and the corresponding zero field dataset
\begin{equation}\label{delta-MM-pm}
	\begin{split}
		\delta\mathbf{M}^{\pm}
		&=
		\mathbf{M}^\pm_{\text{\tiny{OHE}}}
		\hspace{-2pt}-\hspace{-2pt}
		\mathbf{M}_0\\
		&=
		\Delta\mathbf{M}(\eps_{_{\hspace{-1pt} \mathbf{B}=0}},\eps_{_{\hspace{-2pt}\pm \mathbf{B}}})\;,
	\end{split}
\end{equation}
where $\mathbf{M}_0=\mathbf{M}(\eps_{_{\hspace{-1pt}\mathbf{B}=0}})$ is the Mueller matrix of the zero field experiment and $\Delta\mathbf{M}(\eps_{_{\hspace{-1pt} \mathbf{B}=0}},\eps_{_{\hspace{-2pt}\pm \mathbf{B}}})$ is the magnetic field induced change of the Mueller matrix. This form of presentation is in particular advantageous in case the magnetic field causes only small changes in the Mueller matrix, 
 and provides improved sensitivity to magnetic field dependent model parameters during data analysis. Another form of presentation for derived OHE data is
\begin{equation}\label{delta-MM-p-m}
	\begin{split}
		\delta\mathbf{M}^{+}\pm\delta\mathbf{M}^{-}
		\hspace{4pt}
		=
		\hspace{10pt}
		&\Delta\mathbf{M}(\eps_{_{\hspace{-1pt} \mathbf{B}=0}},\eps_{_{\hspace{-2pt}+\mathbf{B}}})\\
		\pm
		&\Delta\mathbf{M}(\eps_{_{\hspace{-1pt} \mathbf{B}=0}},\eps_{_{\hspace{-2pt}-\mathbf{B}}})\;,
	\end{split}
\end{equation}
which can be used to inspect symmetry properties of magneto-optic Mueller matrix data, and can help to improve the sensitivity to magnetic field dependent model parameters during data analysis. 

\subsubsection{Mueller matrix data acquisition (GE)}\label{sec:data acquisition}
Rotating element ellipsometers can be classified into two categories:\cite{Fujiwara_2007} (i) rotating analyzer ellipsometers (RAE)\cite{Hauge1980108,CollinsRSI61_1990,SchubertJOSAA13_1996,Fujiwara_2007} in polarizer-sample-rotating-analyzer ($PSA_R$) or rotating-polarizer-sample-analyzer ($P_RSA$) configuration, capable to measure the upper left $3\times 3$ block of the Mueller matrix; (ii) rotating compensators ellipsometers (RCE)\cite{hauge1975rotating,hauge1976generalized,Aspnes:76,Hauge:78,Hauge1980108,Fujiwara_2007} in polarizer-sample-rotating-compensator-analyzer ($PSC_RA$) or polarizer-rotating-compensator-sample-analyzer ($PC_RSA$) configuration, capable to measure the upper left $3\times 4$ or $4\times 3$ block of the Mueller matrix, respectively.

The Mueller matrices of a polarizer~$\mathbf{P}$, analyzer~$\mathbf{A}$, compensator~$\mathbf{C}(\delta)$ with phase shift $\delta$, coordinate rotation along beam path~$\mathbf{R}\left(\theta\right)$ by an angle $\theta$, and of the sample~$\mathbf{M}$ are given by
\begin{equation}
\begin{aligned}
	&\mathbf{P}=\mathbf{A}\hspace{-10pt}&=\frac{1}{2}
	&\begin{bmatrix}
		\hspace{8pt}1&\hspace{17pt}1&\hspace{17pt}0&\hspace{18pt}0&\hspace{5pt}{}\\
		\hspace{8pt}1&\hspace{17pt}1&\hspace{17pt}0&\hspace{18pt}0&\hspace{5pt}{}\\
		\hspace{8pt}0&\hspace{17pt}0&\hspace{17pt}0&\hspace{18pt}0&\hspace{5pt}{}\\
		\hspace{8pt}0&\hspace{17pt}0&\hspace{17pt}0&\hspace{18pt}0&\hspace{5pt}{}
	\end{bmatrix}\;,\\
	&\mathbf{C}\left(\delta\right)
	&=\ \ \
	&\begin{bmatrix}
		\hspace{8pt}1&\hspace{17pt}0&\hspace{10pt}0&\hspace{0pt}0\\
		\hspace{8pt}0&\hspace{17pt}1&\hspace{10pt}0&\hspace{0pt}0\\
		\hspace{8pt}0&\hspace{17pt}0&\hspace{8pt}\cos \delta&\hspace{2pt}\scalebox{0.6}[1.0]{\(-\)}\sin \delta\\
		\hspace{8pt}0&\hspace{17pt}0&\hspace{10pt}\sin \delta&\hspace{7pt}\cos \delta
	\end{bmatrix}\;,\\
	&\mathbf{R}\left(\theta\right)
	&=\ \ \
	&\begin{bmatrix}
		\hspace{8pt}1&\hspace{6pt}0							&\hspace{-12pt}0																&\hspace{1pt}0&\hspace{5pt}{} \\
		\hspace{8pt}0&\hspace{4pt}\cos 2\theta_j	&\hspace{-1pt}\sin 2\theta_j									&\hspace{1pt}0&\hspace{5pt}{} \\
		\hspace{8pt}0&\hspace{-2pt}\scalebox{0.6}[1.0]{\(-\)}\hspace{1pt}\sin 2\theta_j				&\hspace{-1pt}\cos 2\theta_			j						&\hspace{1pt}0&\hspace{5pt}{} \\
		\hspace{8pt}0&\hspace{7pt}0							&\hspace{-12pt}0																&\hspace{1pt}1&\hspace{5pt}{} 
	\end{bmatrix}\;,\\
	&\mathbf{M}
	&=\ \ \
	&\begin{bmatrix}
		\hspace{3pt}M_{11}&\hspace{4pt}M_{12}&\hspace{4pt}M_{13}&\hspace{4pt}M_{14}&\hspace{-2pt}{}  \\
		\hspace{3pt}M_{21}&\hspace{4pt}M_{22}&\hspace{4pt}M_{23}&\hspace{4pt}M_{24}&\hspace{-2pt}{}  \\
		\hspace{3pt}M_{31}&\hspace{4pt}M_{22}&\hspace{4pt}M_{33}&\hspace{4pt}M_{34}&\hspace{-2pt}{}  \\
		\hspace{3pt}M_{41}&\hspace{4pt}M_{42}&\hspace{4pt}M_{43}&\hspace{4pt}M_{44}&\hspace{-2pt}{} 
	\end{bmatrix}\;,
\end{aligned}
\end{equation}
respectively. Execution of the matrix multiplication characteristic for the corresponding ellipsometer type\cite{Fujiwara_2007} shows that, due to the rotation of optical elements, the measured intensity at the detector is typically sinusoidal. Fourier analysis of the detector signal provides Fourier coefficients, which are used to determine the Mueller matrix of the sample (see sec.~\ref{sec:OHE-acq}).

\subsection{Data analysis}\label{sec:data analysis}
Ellipsometry is an indirect experimental technique. Therefore, in general, ellipsometric data analysis invokes model calculations to determine physical parameters in dielectric tensors or the thickness of layers, for instance.\cite{JellisonTSF313-314_1998} Sequences of homogeneous layers with smooth and parallel interfaces are assumed in order to calculate the propagation of light through a layered sample, by the 4$\times$4 matrix formalism.\cite{Schubert96,SchubertIRSEBook_2004,Fujiwara_2007} To best match the generated data with experimental results, parameters with significance 
are varied and Mueller matrix data is calculated for all spectral data points, angles of incidence and magnetic fields. During the mean square error (MSE) regression, the generated Mueller matrix data $M_{ij,k}^{\text{G}}$ is compared with the experimental Mueller matrix data $M_{ij,k}^{\text{E}}$ and their match is quantified by the MSE
\begin{equation}
	\text{MSE}=\sqrt{\frac{1}{9S-K}\sum^{4}_{i=1}\sum^{4}_{j=1}\sum^{S}_{k=1}\left(\frac{M_{ij,k}^{\text{E}}-M_{ij,k}^{\text{G}}}{\sigma_{M_{ij,k}^{\text{E}}}}\right)^2}\;,
\end{equation}
where $S$, $K$ and $\sigma_{M_{ij,k}^{\text{E}}}$ denotes the total number of data points, the total number of parameters varied during the non-linear regression process, and the standard deviation of $M_{ij,k}^{\text{E}}$, obtained during the experiment, respectively. For fast convergence of the MSE regression, the Levenberg-Marquardt fitting algorithm is used.\cite{press2007numerical} The MSE regression is interrupted when the decrease in the MSE is smaller than a set threshold and the determined parameters are considered as best model parameters. The sensitivity and possible correlation of the varied parameters is checked and, if necessary, the model is changed and the process is repeated.\cite{HerzingerJAP77_1995,Herzinger95b,Johs93}

\begin{figure*}[!t]
	\centering
  \includegraphics[
	width=0.99\textwidth]{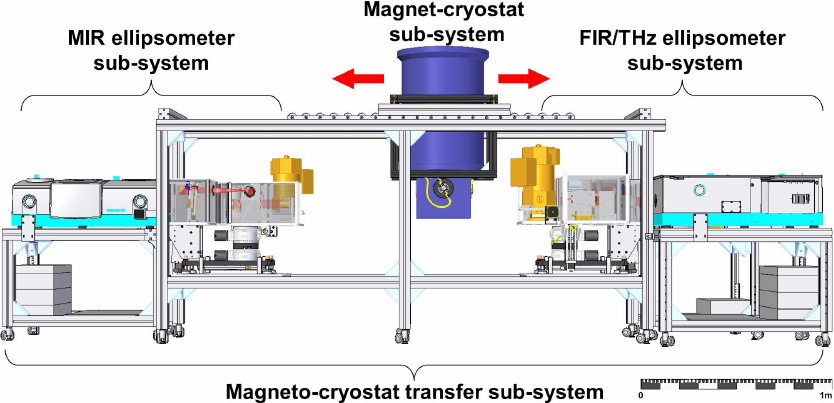}
	\caption{Technical drawing (side view) of the integrated MIR, FIR and THz OHE instrument. The instrument has four sub-systems, (i) the MIR ellipsometer sub-system, (ii) the FIR/THz ellipsometer sub-system, (iii) the magnet-cryostat sub-system, and (iv) the magneto-cryostat transfer sub-system. The magneto-cryostat transfer sub-system holds both ellipsometers and serves as a translation system for the magnet-cryostat sub-system, which can be used with the ellipsometer sub-systems. The total dimensions of the integrated MIR, FIR and THz OHE instrument are 160~cm $\times$ 450~cm $\times$ 115~cm ($h \times w \times d$).}
	\label{fig:CAD-overview}
\end{figure*}
\section{Integrated MIR, FIR and THz OHE instrument}\label{sec:Experimental setup}
\begin{figure}[!t]
	\centering
  \includegraphics[
	width=0.48\textwidth]{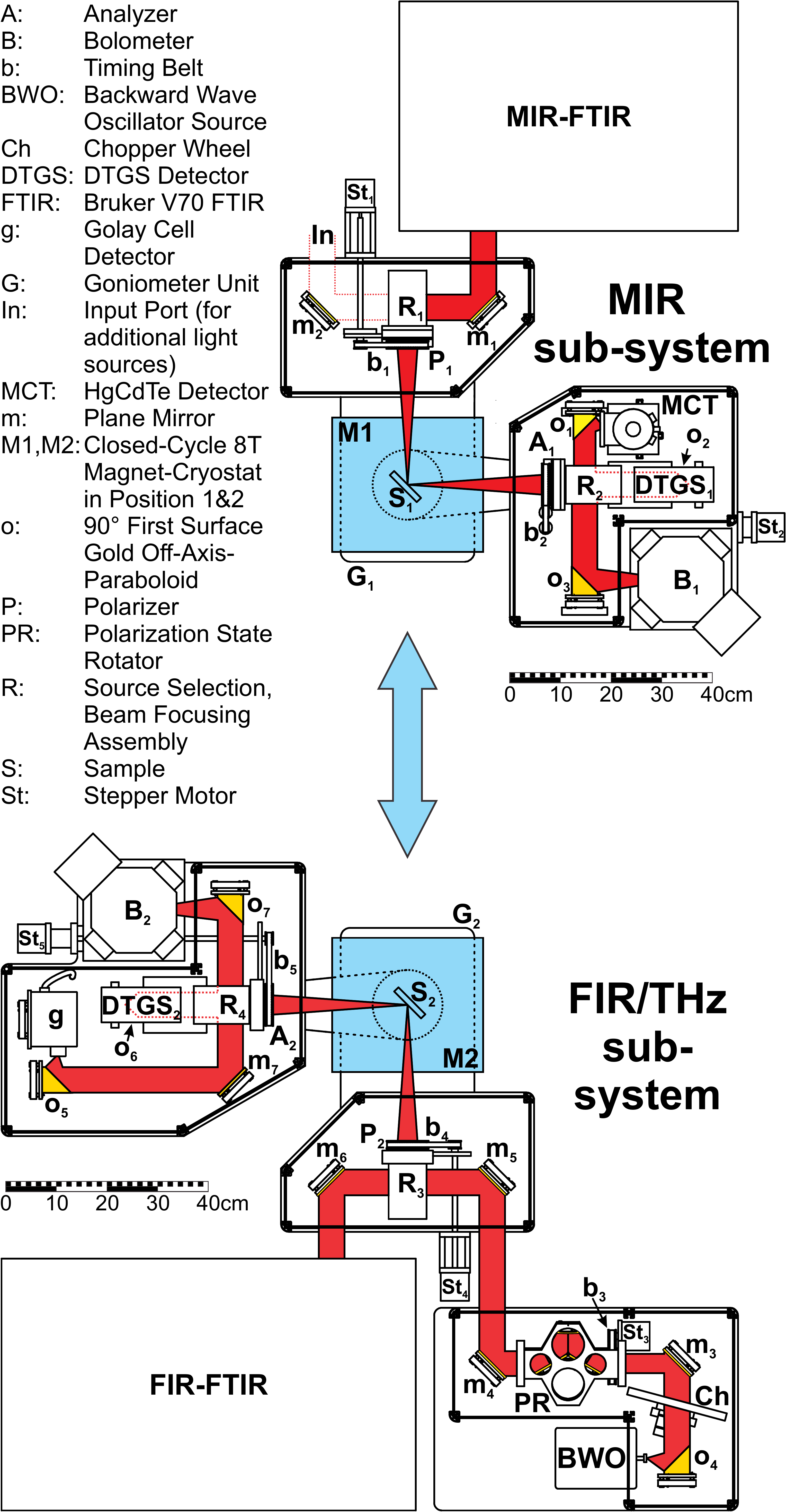}
	\caption{Schematic drawing (top view) of the in-house built, variable angle-of-incidence spectroscopic ellipsometer sub-systems, used for magneto-optic measurements in the wavelength range from 3 to 8000 cm$^{-1}$. In the top part the Fourier-transform-infrared spectroscopy based MIR ellipsometer sub-system is depicted while the lower part shows the combined FIR/THz ellipsometer sub-system. The closed-cycle 8T magnet-cryostat sub-system can be moved between the two ellipsometer sub-systems (M1 or M2) utilizing the magneto-cryostat transfer sub-system (not depicted). }
	\label{fig:2dsetup}
\end{figure}

Figure~\ref{fig:CAD-overview} shows (side view) the integrated MIR, FIR and THz OHE instrument with its four sub-systems: (A) the MIR ellipsometer sub-system, (B) the FIR/THz ellipsometer sub-system, (C) the magneto-cryostat sub-system, and (D) the magneto-cryostat transfer sub-system. In order to utilize the magneto-cryostat sub-system with the MIR or the FIR/THz ellipsometer sub-system the magneto-cryostat transfer sub-system was installed. The integrated MIR, FIR and THz OHE instrument contains multiple light sources and detectors, and covers a spectral range from 3~cm$^{-1}$ to 7000~cm$^{-1}$ (0.1--210~THz or 0.4--870~meV). Both ellipsometer sub-systems can be operated without the magneto-cryostat sub-system, in a variable angle of incidence ellipsometry mode\footnote{The angle of incidence is defined as the angle between the surface normal of the sample and the incoming beam.} ($\Phi_a=30^{\circ}\dots 90^{\circ}$). Figure~\ref{fig:2dsetup} shows a schematic overview (top view) of all major components in the integrated MIR, FIR and THz OHE instrument.


\begin{figure}[tb]
	\centering
  \includegraphics[
	width=0.48\textwidth]{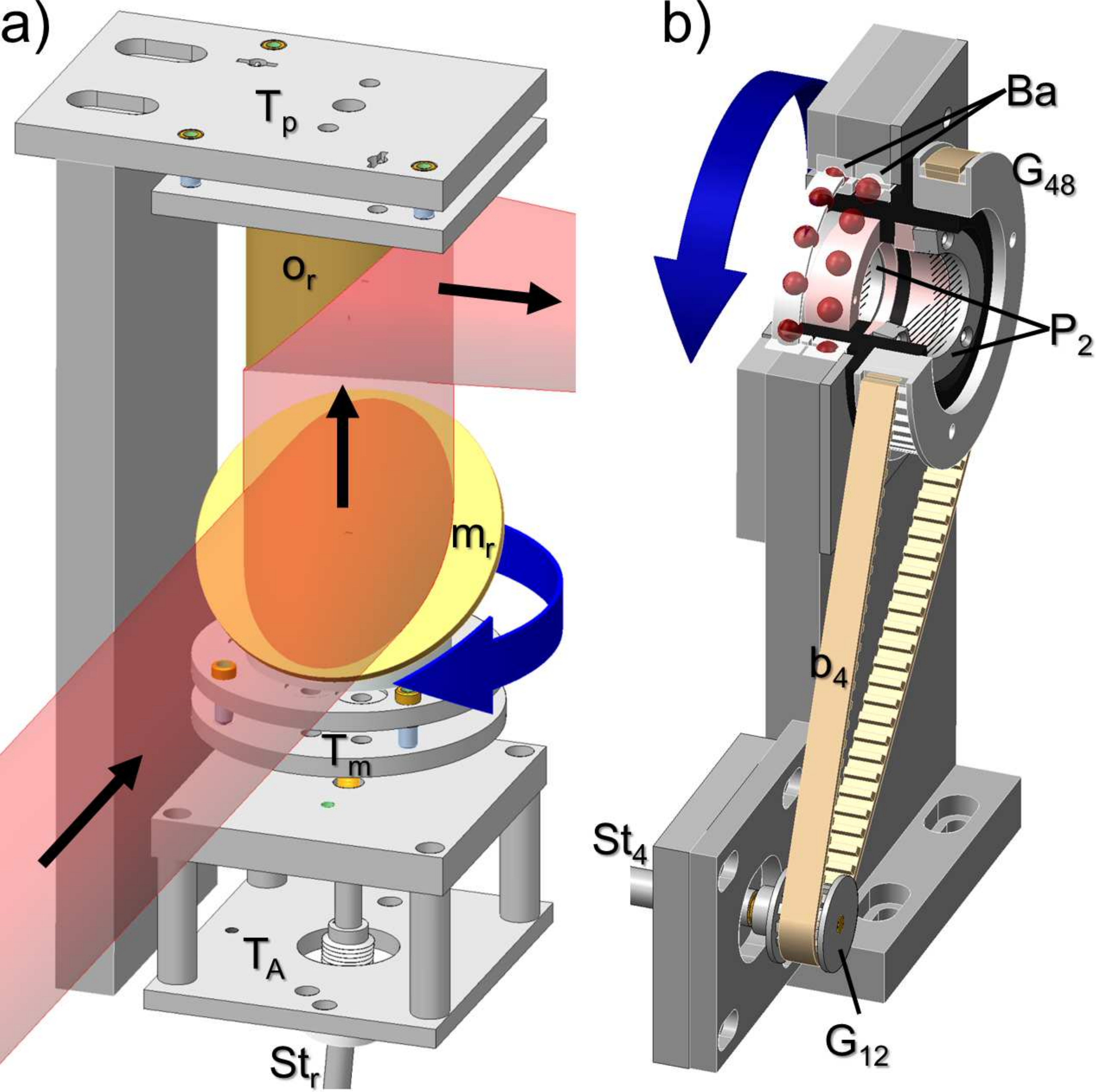}
	\caption{a) Technical drawing of the source selection and beam focusing/beam collimation and detector selection assemblies~(R$_{1-4}$ in Fig.~\ref{fig:2dsetup}). Each source selection and beam focusing/beam collimation and detector selection assembly is composed of a rotatable plane mirror sub-assembly [plane first surface gold mirror~(m$_\text{r}$), opto-mechanic mount for mirror~(T$_\text{m}$), opto-mechanic mount for axis~(T$_\text{A}$), axis with stepper motor and encoder wheel~(St$_\text{r}$)] and a collimating off-axis paraboloid stage sub-assembly [opto-mechanic mount for paraboloid~(T$_\text{p}$), 90$^\circ$ first surface gold off-axis-paraboloid~(o$_\text{r}$)]. b) Technical drawing of the rotation stage assemblies for polarizers and analyzers (here depicted: rotation stage assembly for polarizer of the FIR/THz ellipsometer sub-system, P$_{2}$-b$_{4}$-St$_{4}$ in Fig.~\ref{fig:2dsetup}). The stages comprise an aluminum frame, two plastic bearings with glass balls~(Ba), a shaft which can hold up to two wire grid polarizers~(P$_\text{2}$), two gears [12 teeth~(G$_{12}$) and 48 teeth~(G$_{48}$)], a Kevlar timing belt~(b$_4$) and an axis leading to a stepper motor with encoder wheel~(St$_4$).}
	\label{fig:beamlift}
\end{figure}
\subsection{MIR ellipsometer sub-system}\label{sec:MIR}
\begin{figure}[tbp]
	\centering
  \includegraphics[
	width=0.48\textwidth]{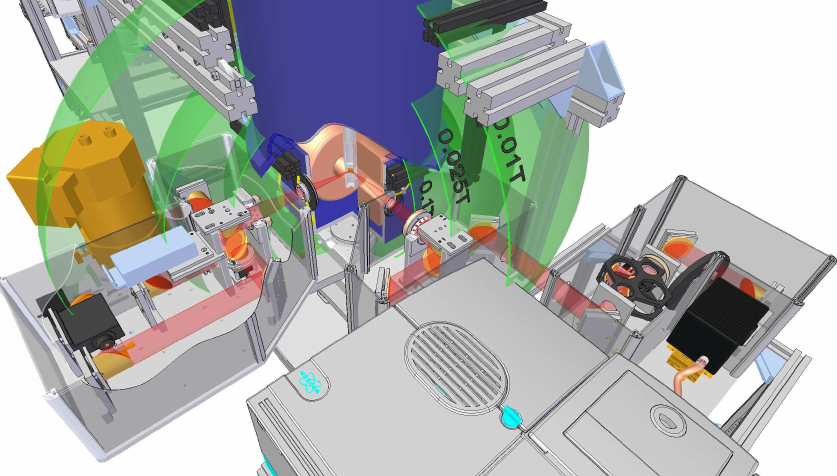}
	\caption{Technical drawing of the FIR/THz ellipsometer sub-system of the integrated MIR, FIR and THz OHE instrument (MIR ellipsometer sub-system not shown) and the magneto-cryostat sub-system. A cutout view of the magneto-cryostat sub-system (blue cylinder, top, center) shows the superconducting magnet coils, the variable temperature inset~(VTI) and the sample. The three cutout prolate ellipsoids (green) represent the spacial positions at which the magnetic stray field is less than 0.1, 0.025 and 0.01~T. The beam path is indicated in red.}
	\label{fig:CAD}
\end{figure}

The upper part of Fig.~\ref{fig:2dsetup} shows a schematic drawing (top view) of the optical configuration of the MIR sub-system of the integrated MIR, FIR and THz OHE instrument. The MIR ellipsometer sub-system is composed of (i) the MIR source unit, (ii) the polarization state preparation unit, (iii) the MIR goniometer unit, and (iv) the polarization state detection unit. To minimize absorption due to water vapor, the complete beam path of the MIR ellipsometer sub-system is purged with dried air.\cite{RSI_Purge_comment} Due to the high magnetic stray-fields (see Fig.~\ref{fig:CAD}), all opto-mechanical components in the polarization state preparation and detection units were designed and manufactured without ferromagnetic materials (with exception of the stepper motors).

The MIR source unit of the MIR ellipsometer sub-system is a Bruker Vertex 70 Fourier-transform-infrared spectrometer (Fig.~\ref{fig:2dsetup}: MIR-FTIR) with a silicon carbide globar light source (spectral range 580--7000~cm$^{-1}$). After being collimated, the light beam passes the interferometer (potassium bromide (KBr) beam splitter), is reflected by a plane mirror, and exits the MIR source unit.

The beam enters the polarization state preparation unit. Inside the polarization state preparation unit the beam passes a beam steering plane mirror assembly~(m$_1$), a source selection and beam focusing assembly~(R$_1$) and a rotation stage assembly~(P$_1$-b$_1$-St$_1$), respectively. The beam steering plane mirror assembly~(m$_1$) is composed of an opto-mechanic mount and a plane first surface gold mirror\cite{RSI_gold_mirror_comment}. The source selection and beam focusing assembly~(R$_1$) comprises two sub-assemblies (detailed drawing: Fig.~\ref{fig:beamlift} a), the rotatable plane mirror sub-assembly\footnote{The rotatable mirror is designed to switch to an alternative input source~(In) (currently unused).} (stepper motor, axis extension, mechanic mount for axis extension, opto-mechanic mount for mirror, plane first surface gold mirror\cite{RSI_gold_mirror_comment}) and the beam focusing off-axis paraboloid stage sub-assembly (opto-mechanic mount for paraboloid, gold surface $90^{\circ}$ off-axis paraboloid with an effective focal length of $f_\text{e}=350$~mm\footnote{The effective focal length $f_\text{e}$ is the distance between the focal point and the center of the off-axis paraboloid mirror. For 90$^{\circ}$ off-axis paraboloids the effective focal length is twice the focal length $f_\text{e}=2f$.}), which focuses the beam onto the sample position. The focused beam then reaches the rotation stage assembly~(P$_1$-b$_1$-St$_1$) which has a nominal angular resolution of 0.045$^\circ$. The rotation stage assembly (detailed drawing: Fig.~\ref{fig:beamlift} b) contains a KRS-5 substrate based wire grid polarizer~(P$_1$), which is mounted in a hollow tube, which is fitted into two polymer bearings with glass balls, which are held by an aluminum block. A 48-tooth polymer gear is mounted to the hollow tube, and is connected via a Kevlar timing belt~(b$_1$) to a 12-tooth polymer gear on a stainless steel shaft (gear ratio:~1:4), leading to the stepper motor~(St$_1$). After passing the rotation stage assembly, the polarized and focused beam leaves the polarization state preparation unit.

The beam is then reflected by, or transmitted through the sample~(S$_1$). The sample can be mounted on a sample holder, attached to the MIR goniometer unit~(G$_1$) (commercially available 2-circle goniometer 415, Huber Diffraktionstechnik), or inside the magneto-cryostat sub-system~(M1). If the magneto-cryostat sub-system is used, reflection type measurements can only be conducted at a $\Phi_a=45^{\circ}$ angle of incidence. A detailed description of the magneto-cryostat sub-system, its sample mount, and the optical window configuration is given in section~(\ref{sec:Magnet}).

The beam then enters the polarization state detection unit, which is mounted to the rotatable arm of the MIR goniometer unit~(G$_1$). The polarization state detection unit contains a rotation stage assembly~(A$_1$-b$_2$-St$_2$), a beam collimation and detector selection assembly~(R$_2$), and three beam focusing/detection assemblies~(o$_1$-MCT, o$_2$-DTGS$_1$ and o$_3$-B$_1$). The rotation stage assembly~(A$_1$-b$_2$-St$_2$) is equivalent to the one in the polarization state preparation unit~(P$_1$-b$_1$-St$_1$, Fig.~\ref{fig:beamlift} b), but in the polarization state detection unit the KRS-5 substrate based wire grid polarizer serves as the analyzer~(A$_1$) of the MIR ellipsometer sub-system. The beam collimation and detector selection assembly~(R$_2$) is composed of two sub-assemblies: the rotatable plane mirror sub-assembly and the collimating off-axis paraboloid stage sub-assembly (gold surface $90^{\circ}$ off-axis paraboloid, $f_\text{e}=350$~mm). Both sub-assemblies are equivalent to the source selection and beam focusing assembly in the polarization state preparation unit (Fig.~\ref{fig:beamlift} a), but are used in reverse order to first collimate and then redirect the beam to one of the three beam focusing/detection assemblies using the rotatable mirror\cite{RSI_gold_mirror_comment}. Each beam focusing/detection assembly contains a beam focusing off-axis paraboloid stage sub-assembly and a detector sub-assembly. All detector sub-assemblies contain an opto-mechanic mount and a detector. The beam focusing off-axis paraboloid stage sub-assemblies~(o$_\text{1-3}$) are composed of an opto-mechanic mount and a 90$^{\circ}$ off-axis paraboloid with an uncoated gold surface. The focal lengths of the off-axis paraboloids are matched to the corresponding detector. The off-axis paraboloids~(o$_\text{1}$, o$_\text{2}$) for the liquid nitrogen cooled HgCdTe-detector sub-assembly~(MCT) and the pyroelectric, solid state deuterated triglycine sulfate detector sub-assembly~(DTGS$_\text{1}$), each have an effective focal length of 38~mm (1.5~in). The off-axis paraboloid~o$_\text{3}$ for the liquid helium cooled bolometer detector sub-assembly~(B$_\text{1}$)\cite{Pankratov78,Pankratov83} has an effective focal length of 190.5~mm (7.5~in). The signal of the used detector is fed back into the MIR-FTIR-spectrometer to record interferograms. For more information on the data acquisition and processing see section~\ref{sec:data analysis}.

\subsection{FIR/THz ellipsometer sub-system}\label{sec:FIR-THz}
The lower part of Fig.~\ref{fig:2dsetup} shows a schematic drawing (top view) of the optical configuration of the FIR/THz ellipsometer sub-system of the integrated MIR, FIR and THz OHE instrument. The FIR/THz ellipsometer sub-system can be divided in five units: (i) the FIR source unit, (ii) the THz source unit, (iii) the polarization state preparation unit, (iv) the FIR/THz goniometer unit, and (v) the polarization state detection unit. For measurements in the FIR spectral range, the FIR/THz ellipsometer sub-system is operated in analyzer-step mode,\cite{Roseler90} while for measurements in the THz spectral range the FIR/THz ellipsometer sub-system is operated in continuously rotating analyzer mode. To minimize absorption due to water vapor, the complete beam path of the FIR/THz ellipsometer sub-system can be purged with dried air.\cite{RSI_Purge_comment} Due to the high magnetic stray-fields (see Fig.~\ref{fig:CAD}), all opto-mechanical components in the THz source unit, polarization state preparation unit and polarization state detection unit were designed and manufactured without ferromagnetic materials (with exception of the stepper motors and the THz source).

The FIR source unit of the FIR/THz ellipsometer sub-system is a Bruker Vertex V-70 FTIR-spectrometer~(Fig.~\ref{fig:2dsetup}: FIR-FTIR). The spectrometer is equipped with a silicon beam splitter but otherwise identical to the spectrometer used as MIR source unit in the MIR ellipsometer sub-system.\\

\begin{figure}[tbp]
	\centering
  \includegraphics[
	width=0.48\textwidth]{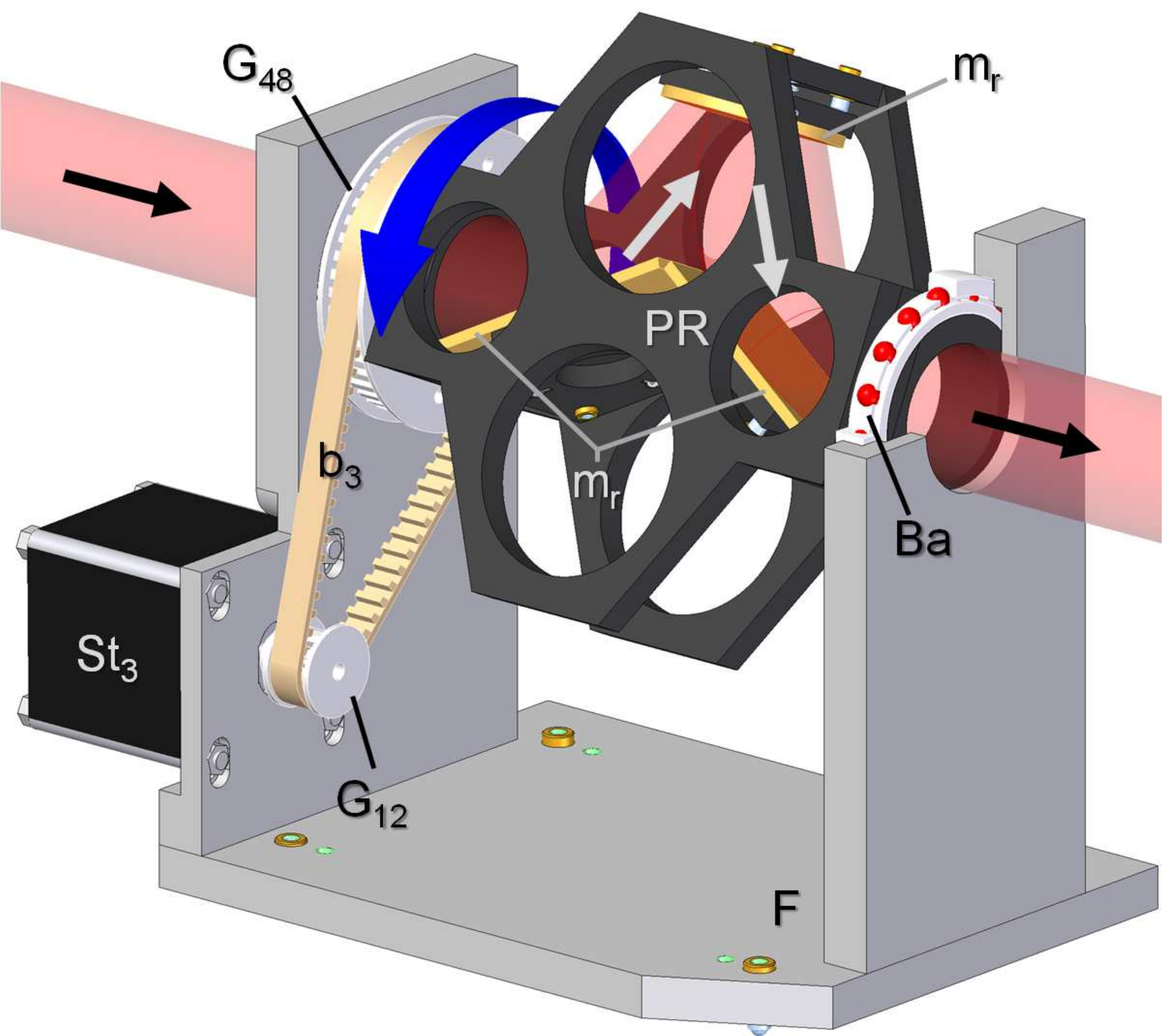}
	\caption{Technical drawing of the polarization state rotator assembly, used in the THz source unit of the FIR/THz ellipsometer sub-system to pre-align the polarization direction of the linearly polarized THz-beam with the polarizing axis of the polarizer in the polarization state preparation unit. The odd-bounce image rotation system\cite{Herzinger_2004} is composed of a frame with an opto-mechanic alumina mount~(F), a stepper motor with encoder wheel~(St$_3$), a 12 teeth gear~(G$_{12}$), connected by a Kevlar timing belt~(b$_3$) to a 48 teeth gear~(G$_{48}$). The 48 teeth gear is glued to a rotatable PEEK cage~(PR) which comprises 3 opto-mechanic mounts with plane first surface gold mirrors~(m$_\text{r}$)\cite{RSI_gold_mirror_comment}, and is mounted into the alumina mount~(F) by two plastic bearings with glass balls~(Ba).}
	\label{fig:PR}
\end{figure}

The THz source unit comprises five assemblies: the THz source and THz-beam collimation assembly~(Fig.~\ref{fig:2dsetup}: BWO-o$_\text{4}$), the optical chopper assembly~(Ch), the beam steering plane mirror assembly~(m$_\text{3}$), polarization state rotator assembly~(PR-b$_\text{3}$-St$_\text{3}$), and the beam steering plane mirror assembly~(m$_\text{4}$). The THz source and THz-beam collimation assembly contains a THz source sub-assembly and a THz-beam collimating off-axis paraboloid stage sub-assembly. The THz source sub-assembly is composed of an opto-mechanic mount and the backward wave oscillator~(BWO) THz source (Microtech). The BWO source emits THz radiation with a high brilliance (bandwidth $\sim$2~MHz) and a high output power ($\sim$0.1--0.01~W). The THz radiation is almost perfectly linearly polarized and the orientation of the polarization is fixed in space. The base frequency range of the BWO is 107--177~GHz, which can be converted to higher frequency bands using GaAs Schottky diode frequency multipliers. The spectral range accessible by the BWO can be expanded to 220--350~GHz ($\times 2$ multiplier), 330--525~GHz ($\times 3$ multiplier), 650--1040~GHz ($\times 2$ and $\times 3$ multiplier) and 980--1580~GHz (double $\times 3$ multiplier). For further details on the BWO based THz source and THz ellipsometry are given in Ref.~\onlinecite{HofmannRSI81_2010} and references therein. The THz radiation from the BWO is collimated by the THz-beam collimating off-axis paraboloid stage sub-assembly, composed of an opto-mechanic mount and a 90$^{\circ}$ off-axis paraboloid~(o$_\text{4}$) with an uncoated gold surface and an effective focal length $f_e=60$~mm. The THz-beam then reaches the optical chopper assembly~(Ch), which contains an opto-mechanic mount and a 3 bladed optical chopper, driven by a linear motor. The 3 bladed optical chopper is rotated with a frequency of $f_c=3.8$~Hz, resulting in a optical chopping frequency of $f_o=11.4$~Hz, which is close to the optimal frequency response of the Golay cell detector ($f_\text{opt}\sim$~12--15~Hz). After interaction with THz-beam steering plane mirror assembly~(m$_\text{3}$) (opto-mechanic mount and plane first surface gold mirror\cite{RSI_gold_mirror_comment}), the THz-beam is redirected to the polarization state rotator assembly~(PR). The polarization state rotator assembly (Fig.~\ref{fig:PR}) is an odd-bounce image rotation system.\cite{Herzinger_2004} The polarization state rotator is designed to rotate the polarization state of an incoming electromagnetic beam azimuthally in a non-deviating, non-displacing fashion (with respect to the incoming electromagnetic beam direction), and is used to pre-align the polarization direction of the THz-beam with the polarizing axis of the wire-grid polarizer in the polarization state preparation unit.\footnote{The stepper motors of the rotation stage in the polarization state preparation unit of the FIR/THz ellipsometer sub-system and the polarization state rotator assembly are operated in tandem mode.} The polarization state rotator assembly is composed of a stepper motor~(St$_3$) with a 12-tooth polymer gear, which is connected to a 48-tooth polymer gear (gear ratio:~1:4) via a Kevlar timing belt~(b$_3$), rotating a PEEK cage~(PR) which contains three opto-mechanic mounts with plane first surface gold mirrors\cite{RSI_gold_mirror_comment} (rotation axis parallel to the incoming and outgoing THz-beam). After reflection on a THz-beam steering plane mirror assembly~(m$_\text{4}$) (opto-mechanic mount and plane first surface gold mirror\cite{RSI_gold_mirror_comment}), the THz-beam leaves the THz source unit.\\

The polarization state preparation unit contains four assemblies: a FIR-beam steering plane mirror assembly~(m$_\text{6}$), a THz-beam steering plane mirror assembly~(m$_\text{5}$), a source selection and beam focusing assembly~(R$_\text{3}$), and  a rotation stage assembly~(P$_\text{2}$-b$_\text{4}$-St$_\text{4}$). The FIR- and THz-beam steering plane mirror assemblies~(m$_\text{6,5}$) are identical and both are composed of an opto-mechanic mount and a plane first surface gold mirror.\cite{RSI_gold_mirror_comment} The plane first surface gold mirrors redirect, depending on the spectral range the FIR/THz ellipsometer sub-system is operated in, the FIR- or THz-beam to the source selection and beam focusing assembly. The source selection and beam focusing assembly~(R$_\text{3}$) comprises two sub-assemblies, the rotatable plane mirror sub-assembly and the beam focusing off-axis paraboloid stage sub-assembly, which are equivalent to those in the source selection and beam focusing assembly in the polarization state preparation unit of the MIR ellipsometer sub-system (Fig.~\ref{fig:beamlift} a). Depending on the orientation of the plane first surface gold mirror\cite{RSI_gold_mirror_comment} in the rotatable plane mirror sub-assembly, either the FIR- or THz beam is directed to beam focusing off-axis paraboloid stage sub-assembly (gold surface $90^{\circ}$ off-axis paraboloid, $f_\text{e}=350$~mm). The focused beam is then routed through the rotation stage assembly, which contains two polyethylene substrate based wire-grid polarizers~(P$_\text{2}$), but is otherwise identical to the rotation stage assembly in the polarization state preparation unit of the MIR ellipsometer sub-system (Fig.~\ref{fig:beamlift} b). The beam then leaves the polarization state preparation unit.

The beam is then reflected by, or transmitted through the sample~(S$_\text{2}$). The sample can be mounted on a sample holder, attached to the FIR/THz goniometer unit~(G$_\text{2}$) (commercially available, 2-circle goniometer 415, Huber Diffraktionstechnik), or inside the magneto-cryostat sub-system~(M2). If the magneto-cryostat sub-system is used, reflection type measurements can only be conducted at $\Phi_a=45^{\circ}$ angle of incidence. A detailed description of the magneto-cryostat sub-system, its sample mount, and the optical window configuration is given in section~\ref{sec:Magnet}.

The beam then enters the polarization state detection unit, which comprises a rotation stage assembly~(A$_\text{2}$-b$_\text{5}$-St$_\text{5}$), a beam collimation and detector selection assembly~(R$_\text{4}$), and three beam focusing/detection assemblies~(m$_\text{7}$-o$_\text{5}$-g, o$_\text{6}$-DTGS$_\text{2}$ and o$_\text{7}$-B$_\text{2}$). The beam is routed through the rotation stage assembly for the analyzer of the FIR/THz ellipsometer sub-system, which contains two polyethylene substrate based wire-grid polarizers~(A$_\text{2}$), but is otherwise identical to the rotation stage assembly in the polarization state detection unit of the MIR ellipsometer sub-system (Fig.~\ref{fig:beamlift} b). The rotation stage assembly for the analyzer of the FIR/THz ellipsometer sub-system can be operated in step mode for FIR measurements\cite{Roseler90} or in continuous rotation mode for THz measurements. The beam is then collimated (gold surface $90^{\circ}$ off-axis paraboloid, $f_\text{e}=350$~mm) and redirected to the selected detector by the beam collimation and detector selection assembly~(R$_\text{4}$), identical to the beam collimation and detector selection assembly in the polarization state detection unit of the MIR ellipsometer sub-system (Fig.~\ref{fig:beamlift} a). The beam focusing and Golay-cell-detector assembly~(m$_\text{7}$-o$_\text{5}$-g) contains a beam steering plane mirror sub-assembly~(m$_\text{7}$) (opto-mechanic mount, plane first surface gold mirror\cite{RSI_gold_mirror_comment}), a beam focusing off-axis paraboloid stage sub-assembly~(o$_\text{5}$) (opto-mechanic mount, gold surface $90^{\circ}$ off-axis paraboloid, $f_\text{e}=60$~mm), and a Golay-cell detector sub-assembly~(g) (opto-mechanic mount, Golay-cell detector). The beam focusing and DTGS detector assembly~(o$_\text{6}$-DTGS$_\text{2}$) comprises a beam focusing off-axis paraboloid stage sub-assembly~(o$_\text{6}$) (opto-mechanic mount, gold surface $90^{\circ}$ off-axis paraboloid, $f_\text{e}=38$~mm), and a solid state deuterated triglycine sulfate detector sub-assembly~(DTGS$_\text{2}$) (opto-mechanic mount, Bruker Vertex V-70 DTGS detector). Alternatively, the beam focusing and bolometer detector assembly~(o$_\text{7}$-B$_\text{2}$), composed of a beam focusing off-axis paraboloid stage sub-assembly~(o$_\text{7}$) (opto-mechanic mount, gold surface $90^{\circ}$ off-axis paraboloid, $f_\text{e}=190.5$~mm) and the bolometer detector sub-assembly~(B$_\text{2}$) (opto-mechanic mount; commercially available, liquid helium cooled bolometer detector, Infrared Laboratories Inc.), can be used. For THz measurements the bolometer or the golay cell detector can be chosen, while for FIR measurements only the bolometer or the DTGS detector provide frequency responses fast enough to record interferograms.

\subsection{Magneto-cryostat sub-system}\label{sec:Magnet}
The central piece of equipment of the integrated MIR, FIR and THz OHE instrument is the commercially available, superconducting, closed cycle magneto-cryostat sub-system (7T-SpectromagPT, Oxford Instruments) with four optical ports (Fig.~\ref{fig:CAD}). The design of the integrated MIR, FIR and THz OHE instrument allows the usage of the magneto-cryostat sub-system with the MIR and the FIR/THz ellipsometer sub-system by employing the magneto-cryostat transfer sub-system. The magneto-cryostat sub-system can be subdivided into the magnet head, the primary cooling cycle, the secondary cooling cycle, the sample holder and the optical windows.

The magnet head contains two magnet coils, which are mounted around the sample position and are fabricated in a split-coil pair design. In a spherical volume of 10~mm diameter around the sample, magnetic fields up to $B=8$~T with an inhomogeneity of less than 0.3~\% can be achieved. The magnetic field can be reversed and points towards one of the optical windows. Therefore, for reflection type OHE measurements, the magnetic field lies within the plane of incidence and forms an angle of $45^{\circ}$ with the sample normal. This leads to a magnetic field component $B_c=|\mathbf{B}|/\sqrt{2}$ perpendicular to the sample surface.

The primary cooling cycle comprises a pulse tube cooler (SRP-082, SHI Cryogenics), high pressure helium lines and a helium compressor (F-70, Sumitomo Heavy Industries). Ultra-high-purity helium~(UHP-He) gas at high pressure is provided by the helium compressor, and guided by a high pressure helium line to the pulse tube cooler. The pulse tube cooler is thermally coupled to the superconducting magnet coils and allows to cool the magnet coils to temperatures of $T\approx 3.1$~K. The pulse tube cooler also pre-cools the UHP-He in the secondary cooling cycle. The UHP-He gas is then guided back to the helium compressor by a high pressure helium line, and is reused.

\begin{figure}[!t]
	\centering
  \includegraphics[
	width=0.48\textwidth]{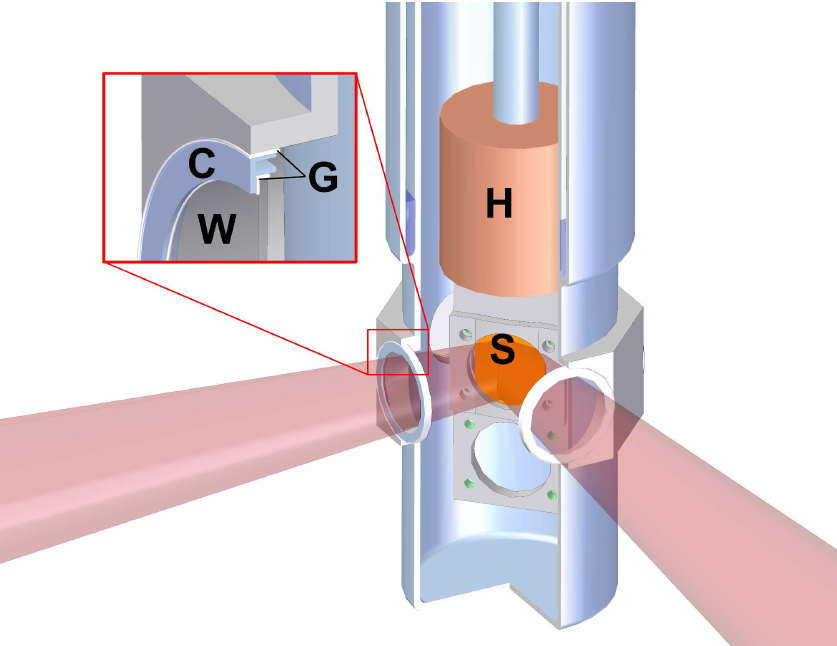}
	\caption{Technical drawing of the sample~(S) and its holder, including the sample heater, capsuled in a copper block~(H). 
	The sample is thermally coupled to the variable temperature inset~(VTI) by a static exchange gas~(UHP-He) surrounding the sample. The inner, 0.35~mm thick, diamond windows~(W) are wedged and were glued to their stainless steel window frames~(C) by a two component epoxy~(G).}
	\label{fig:sampleholder}
\end{figure}
The secondary cooling cycle uses UHP-He gas to cool the sample. The UHP-He gas in the closed cycle, is circulated by an oil-free, dry scroll pump (XDS-10, Edwards). When the sample is at base temperature, the UHP-He gas leaves the outlet of the scroll pump at a pressure of $\sim$~0.5~bar, and is pumped through a zeolite and a liquid nitrogen trap, in order to extract possible contaminants leaking into the closed cycle. The UHP-He gas then passes a distiller spiral, which is thermally coupled to the pulse tube cooler (primary cooling cycle), and condenses the UHP-He gas into liquid helium (LHe). The LHe flow is controlled by a needle valve, and the LHe is then injected into a heat exchanger. The heat exchanger is attached to the double-walled, hollow variable temperature inset (VTI) cryostat (Fig.~\ref{fig:sampleholder}). Finally, the scroll pump reduces the gas pressure above the LHe to 1.5-3~mbar and thereby cools the VTI to a minimal temperature of $T=1.4$~K. The sample is thermally coupled to the VTI by a static exchange gas (UHP-He). A resistive heater allows to warm the sample up to room temperature, without bringing the temperature of the superconducting magnet coils above the critical temperature.

The sample holder of the magneto-cryostat sub-system (Fig.~\ref{fig:sampleholder}) can hold up to two samples at a time. If two samples are mounted the optimal sample size is \mbox{$0.5\times 12\times 12$~mm$^3$}, while the maximum sample size is \mbox{$1\times 30\times 30$~mm$^3$} (only one sample can be mounted). The sample position is adjustable in the vertical direction ($\pm15$~mm) and rotationally around the vertical axis of  the VTI by 360$^{\circ}$ (angle of incidence alignment). All other degrees of freedom necessary for sample alignment (linear motion, rotation, tip/tilt) can only be accessed by moving the sample together with the magneto-cryostat sub-system. The magneto-cryostat sub-system can be moved parallel to the incoming beam, using the magneto-cryostat transfer sub-system. The alignment perpendicular to the incoming beam and the rotational alignment of the magneto-cryostat sub-system is achieved by sliding the magneto-cryostat sub-system in the magneto-cryostat holding frame of the magneto-cryostat transfer sub-system (see sec.~\ref{sec:Magnet_transfer}). Four screws in the same magneto-cryostat holding frame of the magneto-cryostat transfer sub-system are used for tip/tilt alignment. After successful alignment, in order to minimize motion, the magneto-cryostat sub-system is clamped to the magneto-cryostat holding frame of the magneto-cryostat transfer sub-system.

When the light beam is routed through the magnet head, it passes a set of exterior and interior optical windows. For measurements in the FIR/THz spectral range, the exterior optical windows are made of 0.27~mm thick homo-polypropylene films, while for the MIR spectral range potassium bromide (KBr) windows are used. All four exterior windows are purged on the exterior side with dried air, to prevent condensation of moisture from the ambient air. The exterior optical windows of the magnet can be exchanged and arranged for both transmission- and reflection-type measurements. The latter window configuration allows reflection-type measurements over the full spectral range of the OHE instrument, by simply moving the magneto-cryostat between the ellipsometer sub-systems without warming up the superconducting coils or the sample. The interior four optical windows on the VTI are made of polished diamond, grown by chemical vapor deposition (CVD), with a thickness of 0.35~mm, a diameter of 14~mm and an average surface roughness of $Ra\leq 15$~nm (arithmetic average). To reduce Fabry-P\'erot interferences in the interior optical windows the interior windows were wedged by an angle of $\sim$0.5$^{\circ}$. The interior windows were glued into stainless steel window frames by a cryogenic two component epoxy, leaving a clear aperture of $\sim$12~mm (see cutout in Fig.~\ref{fig:sampleholder}).

In general, optical windows affect experimental ellipsometry data, especially if the optical window material is birefringent. Therefore all optical windows were characterized by transmission GE on a commercial MIR ellipsometer (J.A.Woollam Co., Inc.). No birefringence was observed in the MIR spectral range. Nevertheless, through mounting of the optical windows, and in particular through stress due to the vacuum in the magneto-cryostat sub-system, strain induced birefringence cannot be completely excluded and therefore window effects are included in the analysis of the OHE data (see sec.~\ref{sec:OHE-analysis}).

\subsection{Magneto-cryostat transfer sub-system}\label{sec:Magnet_transfer}
The magneto-cryostat transfer sub-system (Fig.~\ref{fig:CAD-overview}) contains the magneto-cryostat transfer frame, the MIR source unit frame and the FIR/THz source unit frame. The magneto-cryostat transfer frame comprises the magneto-cryostat transfer assembly, the MIR goniometer unit platform assembly and FIR/THz goniometer unit platform assembly. The magneto-cryostat transfer assembly is built from alumina extrusions. The grooves of the alumina extrusions are used as guides for a Teflon roll based rail sub-assembly. On top of the Teflon roll based rail sub-assembly the magneto-cryostat holding frame (black alumina extrusions in Fig.~\ref{fig:CAD-overview} and \ref{fig:CAD}) is mounted, which is used for alignment purposes (see sec.~\ref{sec:Magnet}). The MIR and FIR/THz goniometer unit platform assemblies are mounted into the magneto-cryostat transfer frame and are platforms for the MIR and FIR/THz goniometer unit of the ellipsometer sub-systems. The MIR and FIR/THz source unit frames are built from alumina extrusions, equipped with a 19-inch rack mounting system, and provide platforms for the source units of the MIR and FIR/THz ellipsometer sub-systems, respectively.

\begin{figure*}[htbp]
	\centering
  \includegraphics[
	width=0.88\textwidth]{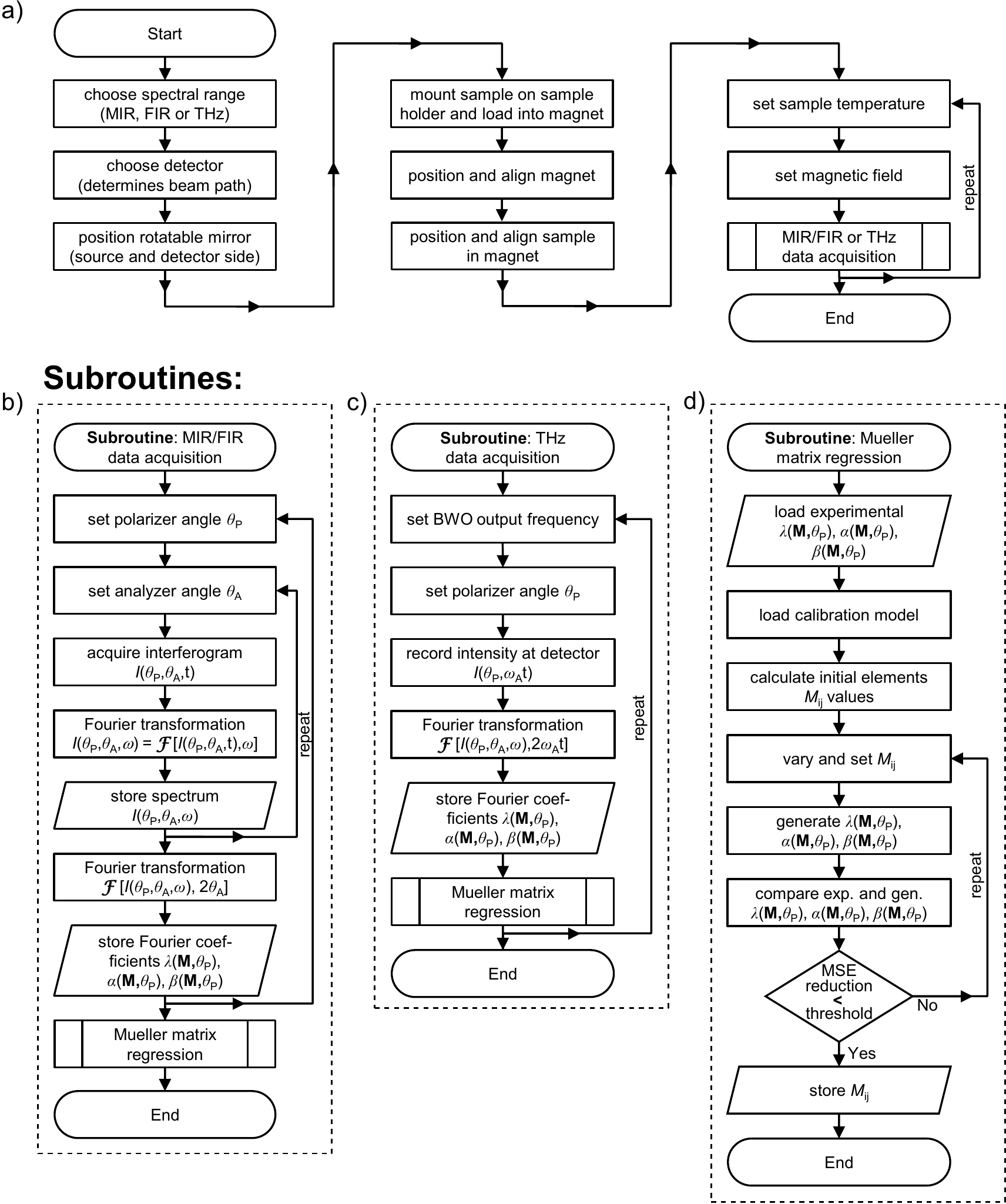}
	\caption{Flowchart of the OHE-data acquisition process. a) Shows the OHE-data acquisition process for all spectroscopic ellipsometers sub-systems. b) Summarizes the subroutine for MIR/FIR data acquisition, c) for THz data acquisition and d) summarizes the Mueller matrix regression process (subroutine of the MIR/FIR and THz data acquisition).}
	\label{fig:flow_aqu}
\end{figure*}
\section{OHE data acquisition and analysis}\label{sec:ACQ-ANAL}
\subsection{OHE data acquisition}\label{sec:OHE-acq}
All spectroscopic ellipsometer sub-systems of the integrated MIR, FIR and THz OHE instrument are operated in a $PSA_R$ configuration, capable to measure the upper left $3\times 3$ block of the Mueller matrix. Note that this does not affect the ability to determine certain sample properties related to anisotropy, which can be obtained from the off-diagonal-block elements $M_{13}$, $M_{23}$, $M_{31}$ and $M_{32}$. For rotating-analyzer based ellipsometers with lossless and ideal polarizing optical elements, the stokes vector of a beam of light at the detector $\mathbf{I}^{\text{D}}$ can be described within the Mueller matrix formalism by\cite{Fujiwara_2007}
\begin{equation}
	\label{eqn:MatMulElli}
		\mathbf{I}^{\text{D}}
		\hspace{-2pt}=\hspace{-2pt}
		\mathbf{R}(-\theta_\text{A})
\hspace{2pt}
		\mathbf{A}
\hspace{2pt}
		\mathbf{R}(\theta_\text{A})
\hspace{2pt}
		\mathbf{M}
\hspace{2pt}
		\mathbf{R}(-\theta_\text{P})
\hspace{2pt}
		\mathbf{P}
\hspace{2pt}
		\mathbf{R}(\theta_\text{P})
\hspace{2pt}
		\mathbf{I}^{\text{S}}\;,
\end{equation}
where $\mathbf{I}^{\text{S}}$ is the stokes vector of the light leaving the source, and $\mathbf{P}$, $\mathbf{A}$, $\mathbf{R}\left(\theta_j\right)$ and $\mathbf{M}$ are the Mueller matrices of a polarizer, analyzer, coordinate rotation by the angle $\theta_j$ (polarizer: $j=\text{P}$; analyzer: $j=\text{A}$) and sample, respectively.

If the analyzer is rotated with a constant angular frequency $\omega_\text{A}$, and both, light source and detector exhibit no polarization dependency, \textit{e.g.} the source emits unpolarized light with intensity $I^{\text{S}}_0$ and the detector is only sensitive to the total intensity $I^{\text{D}}_0$,\footnote{In this case the stokes vectors are given by $\left(\mathbf{I}^{\text{S}}\right)^T=\left(I^{\text{S}}_0,0,0,0\right)$ and $\left(\mathbf{I}^{\text{D}}\right)^T=\left(I^{\text{D}}_0,0,0,0\right)$} the ratio of these quantities is
\begin{equation}\label{eqn:fouriercoef}
	\begin{split}
		\frac{I^{\text{D}}_0}{I^{\text{S}}_0}\hspace{-2pt}
		=
		\hspace{-2pt}
		\frac{1}{4}
		\hspace{-2pt}
		\left[
			\lambda(\mathbf{M}\hspace{-1pt},\hspace{-1pt}\theta_{\hspace{-0pt}\text{P}}\hspace{-2pt})\hspace{-2pt}
		\right.
		&
		+
		\hspace{-2pt}
		\alpha(\mathbf{M}\hspace{-1pt},\hspace{-1pt}\theta_{\hspace{-0pt}\text{P}}\hspace{-2pt})
		\cos (2 \omega_\text{A} t)
		\hspace{-2pt}\\
		&
		+
		\left.
			\hspace{-2pt}
			\beta(\mathbf{M}\hspace{-1pt},\hspace{-1pt}\theta_{\hspace{-0pt}\text{P}}\hspace{-2pt}) 
			\sin(2 \omega_\text{A} t)
		\right]\;,
	\end{split}
\end{equation}
with the time harmonic Fourier coefficients
\begin{equation}\label{eqn:NCSparameter}
\begin{split}
		\lambda (\mathbf{M},\theta_\text{P})
		=&
		M_{11} + M_{12} \cos(2 \theta_\text{P}) + M_{13} \sin(2 \theta_\text{P})\\
		\alpha (\mathbf{M},\theta_\text{P})
		=&
		M_{21} + M_{22} \cos(2 \theta_\text{P}) + M_{23} \sin(2 \theta_\text{P})\\
		\beta (\mathbf{M},\theta_\text{P})
		=&
		M_{31} + M_{32} \cos(2 \theta_\text{P}) + M_{33} \sin(2 \theta_\text{P})\;.
	\end{split}
\end{equation}
Individual Mueller matrix elements $M_{ij}$ are determined from Fourier coefficients measured at different polarizer orientations $\theta_\text{P}$ on the input side (see subroutine Mueller matrix regression in Fig.~\ref{fig:flow_aqu}). The complete sequence of operations executed by the OHE instrument during the GE and OHE data acquisition 
is summarized 
in Fig.~\ref{fig:flow_aqu}.

Since the intensity of light at the detector and the Fourier coefficients (employed to determine the Mueller matrix elements) depend on the absolute angular positions of all rotating optical elements [Eq.~(\ref{eqn:fouriercoef}) and (\ref{eqn:NCSparameter})], the knowledge of these absolute angular positions is crucial for the operation of every ellipsometer. Furthermore, all spectroscopic ellipsometers have to account for non-ideal polarization characteristics of sources and detectors, as well as commonly present non-idealities of MIR and FIR optical elements (such as polarizers), in their calibration routine. The absolute positions of the optical components used in the OHE instrument, with respect to the \textit{p}- and \textit{s}-coordinate system (the \textit{s}-axis is parallel to the goniometer axis, the \textit{p}-axis is perpendicular to the beam), as well as the non-idealities of the optical elements are carefully calibrated prior to the OHE data acquisition. All ellipsometers sub-systems are calibrated prior to the experiments.

\begin{figure}[tb]
	\centering
  \includegraphics[
	width=0.32\textwidth]{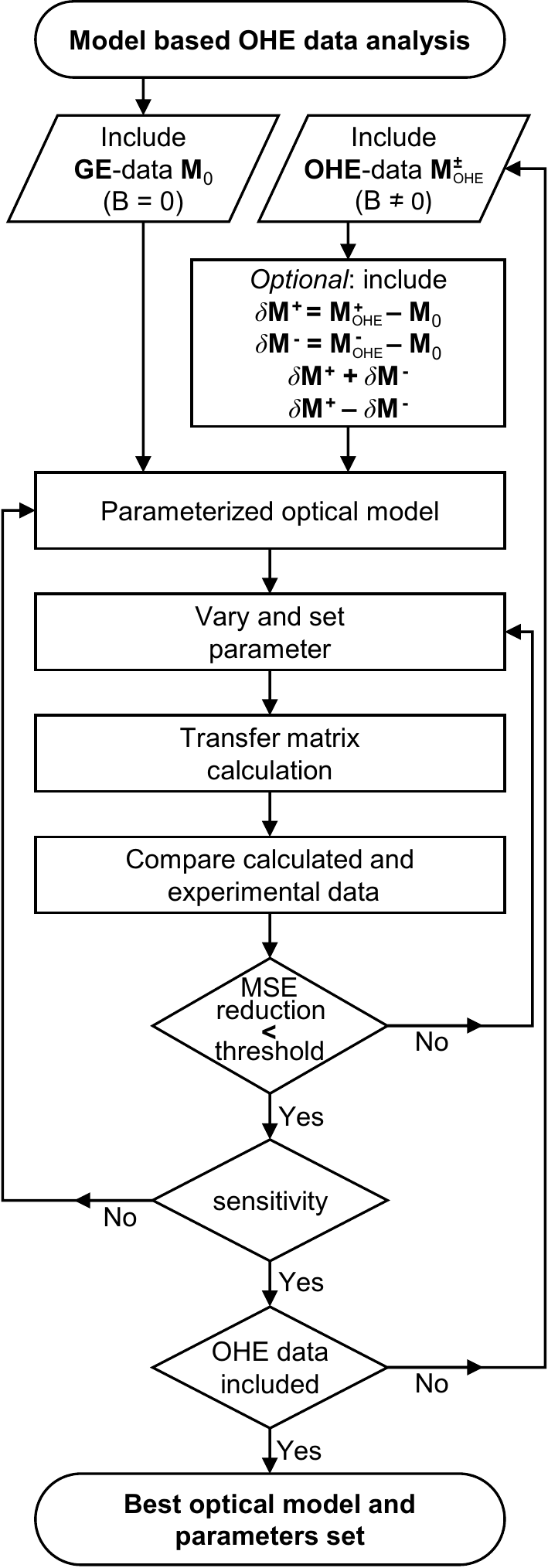}
	\caption{Flowchart of model based analysis of OHE-data. In a first step only experimental data for $B=0$~T is analyzed and only best-model parameters \textit{independent} on the magnetic field are obtained. Their values are then used as starting parameters in a second step, where the whole OHE dataset is utilized for analysis, and best-model parameters \textit{dependent} on the magnetic field are determined.}
	\label{fig:flow}
\end{figure}
\subsection{OHE data analysis}\label{sec:OHE-analysis}
\begin{figure*}[tbp]
	\centering
  \includegraphics[
	width=0.99\textwidth]{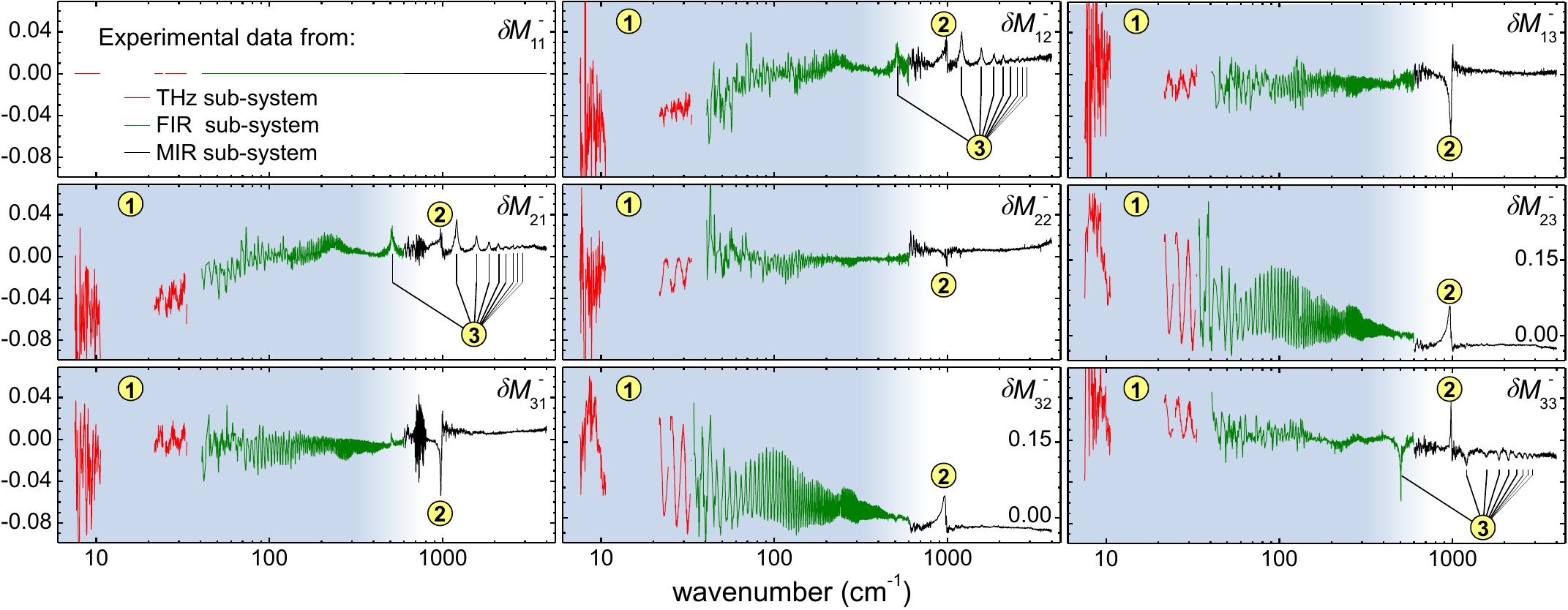}
	\caption{Wide spectral range OHE experiment: derived OHE data $\delta\mathbf{M}^{-}$ from 7.5~cm$^{-1}$ to 4000$^{-1}$ (0.22--120~THz or 0.9--500~meV) for epitaxial graphene on 6\textit{H}-SiC at $B=-4$~T. \large\ding{172}\normalsize: Below 500-600~cm$^{-1}$ the experimental data is dominated by Fabry-P\'erot interferences in the 6\textit{H}-SiC substrate (indicated by the background color). Fabry-P\'erot interferences typically enhanced the sensitivity to the OHE greatly. \large\ding{173}\normalsize: The 6\textit{H}-SiC substrate phonon mode greatly enhances the magnitude of the OHE signal from free charge carriers. \large\ding{174}\normalsize: Single layer graphene inter-Landau-level transitions are only observed in the on-diagonal block elements of the Mueller matrix. Note the different $y$-axis scale for the matrix elements $\delta M_{23}^-$ and $\delta M_{32}^-$.}
	\label{fig:full_range}
\end{figure*}
During OHE data analysis, additional non-idealities, not included in the ellipsometer calibration, introduced by the magneto-cryostat sub-system have to be considered. In particular, three effects are considered: (i) in- and-out-of-plane anisotropy in the optical window\cite{WoPat,Nijs1988}, (ii) changes in the alignment of mirrors and/or off-axis paraboloids in the polarization state preparation and detection units due to magnetic forces on ferromagnetic components (ellipsometric coordinate system change), and (iii) imperfect sample alignment (ellipsometric coordinate system and angle-of-incidence change). In order to account for sample misalignment in the model based data analysis described in section~\ref{sec:data analysis}, the angle of incidence and the sample tilt angle\footnote{The sample tilt angle is defined as the angle by which the sample is rotated along the intersection of the plane off incidence and the sample surface.} are included as model parameters. In order to model the combined non-idealities due to strain induced birefringence in the optical windows and minute changes in the alignment of optical elements on the input (output) side, an additional Mueller matrix $\mathbf{M}_{\text{\tiny{in}}}$ ($\mathbf{M}_{\text{\tiny{out}}}$) was included in the model based data analysis (Sec.~\ref{sec:data analysis}). The Mueller matrices $\mathbf{M}_{\text{\tiny{in}}}$ and $\mathbf{M}_{\text{\tiny{out}}}$ are assumed to have no dispersion (wavelength independent). The best-model Mueller matrix $\mathbf{M}_{\text{\tiny{best}}}$, used for MSE regression as described in section~\ref{sec:data analysis} reads
\begin{equation}
	\mathbf{M}_{\text{\tiny{best}}}
	=
	\mathbf{M}_{\text{\tiny{out}}}
	\mathbf{M}_{\text{\tiny{mod}}}
	\mathbf{M}_{\text{\tiny{in}}}
	\;,
\end{equation}
where $\mathbf{M}_{\text{\tiny{mod}}}$ represents the sample Mueller matrix calculated by the $4\times 4$ matrix formalism.\footnote{The variation of the best-model parameters $M_{ij}$ of the in- and output matrices $\bm{M}_{\text{\tiny{in}}}$ and $\bm{M}_{\text{\tiny{out}}}$ is smaller then 0.02 over the full magnetic field range.}

For the analysis of OHE data, two strategies may be used exploiting either model-free or model-based approaches. Model free analysis provides semi-quantitative results, by studying trends in amplitudes or spectral positions of features in OHE or derived OHE data vs. the magnitude of the magnetic field $|\mathbf{B}|$. The model free analysis can provide insight into the symmetry properties of magneto-optic dielectric tensors,\cite{KuehnePRL111_2013} and  an example for the model free analysis of derived OHE data $\delta\mathbf{M}^{+}$ vs. $|\mathbf{B}|$ is given in section~\ref{sec:Graphene} for the case of epitaxial graphene.

The model-based data analysis approach provides more quantitative parameters than the model-free data analysis approach, and can be  used to determine model parameters such as the free charge carrier concentration or the effective mass parameters. During the data analysis the GE, OHE and derived OHE datasets are analyzed simultaneously to determine physical model parameters of $\eps_{_{\hspace{-1pt} \mathbf{B}=0}}$ and $\eps_{_{\hspace{-2pt}\mathbf{B}}}$ by a single, consistent optical model (Fig.~\ref{fig:flow}).
\begin{itemize}
	\item In the first step, $\mathbf{M}_0=\mathbf{M}(\eps_{_{\hspace{-1pt}\mathbf{B}=0}})$ (GE data obtained at $B=0$~T) is analyzed only. During the analysis, all model parameters \textit{independent} of the magnetic field, such as those describing, for example, the polar lattice resonances or layer thickness are varied until the calculated data match the measured data as closely as possible.
	\item In the second step, OHE data $\mathbf{M}^\pm_{\text{\tiny{OHE}}}$ for $B\neq 0$~T is included in the analysis, and all model parameters \textit{dependent} on the magnetic field, such as, for example, effective mass or inter-Landau-level transition parameters are varied. In addition, derived OHE data $\delta\mathbf{M}^{\pm}$ and/or $\delta\mathbf{M}^{+}\pm\delta\mathbf{M}^{-}$ can be included into the simultaneously analyzed dataset. These derived OHE datasets do not increase the amount of information collected during the experiment, but help (\textit{a}) to visualize magnetic field induced changes, and (\textit{b}) improve the sensitivity to magnetic-field dependent model parameters during OHE data analysis.\footnote{Including derived OHE datasets in the data analysis is in particular useful if magneto-optic effects only lead to a subtle change of the Mueller matrix with respect to the Mueller matrix obtained at $B=0$~T, as for example in case of inter-Landau-level transitions in epitaxial graphene, discussed in section~\ref{sec:Graphene}.}
\end{itemize}
The parameters determined in the first analysis step are used as starting values for the second analysis step and are varied if necessary. Note that an optical model, describing the GE dataset ($B=0$~T) sufficiently well, might not be capable to describe the complete dataset including OHE and/or derived OHE data correctly. If necessary the optical model has to be adjusted, and the data analysis procedure has to be repeated. 
Additional information on the data analysis strategies can be found in Ref.~\onlinecite{HofmannRSI77_2006}.

\section{Results and Discussions}\label{sec:samplesystems}
In this section, experimental data from the integrated MIR, FIR and THz OHE instrument is presented. First, a combined dataset of derived OHE data from the MIR, FIR and THz spectral range from 7.5~cm$^{-1}$ to 4000$^{-1}$ (0.22--120~THz or 0.9--500~meV), and a magnetic field of $|\mathbf{B}|=$4~T, for an epitaxial graphene sample grown on 6\textit{H}-SiC is shown (Fig.~\ref{fig:full_range}). In the spectral range below approximately 500-600~cm$^{-1}$ the experimental data reveals an OHE signal enhanced by Fabry-P\'erot interferences in the 6\textit{H}-SiC. An OHE signal, enhanced by coupling with the 6\textit{H}-SiC phonon mode, is observed near 1000~cm$^{-1}$. Furthermore, between 500 and 4000~cm$^{-1}$ inter-Landau-level transitions in single layer graphene are observed in the on-diagonal block elements of the Mueller matrix.

Then derived OHE data from the individual MIR, FIR and THz spectral ranges of the integrated MIR, FIR and THz OHE instrument, and the corresponding best-model calculated data are shown exemplarily. We present results from OHE experiments on an epitaxial graphene sample grown on 6\textit{H}-SiC, a Te doped n-type GaAs substrate and an AlGaN/GaN high electron mobility transistor structure~(HEMT), representing the MIR, FIR, and THz spectral range of the integrated MIR, FIR and THz OHE instrument, respectively. The selected experimental datasets demonstrate the full spectral, magnetic field and temperature range of the integrated MIR, FIR and THz OHE instrument, as well as analysis strategies. Effects from free charge carriers in bulk and in two dimensional confinement as well as quantum mechanical effects (inter-Landau-level transitions) are observed and discussed.

\subsection{The MIR optical Hall effect --- Graphene}\label{sec:Graphene}
\begin{figure}[tbp]
	\centering
  \includegraphics[
	width=0.48\textwidth]{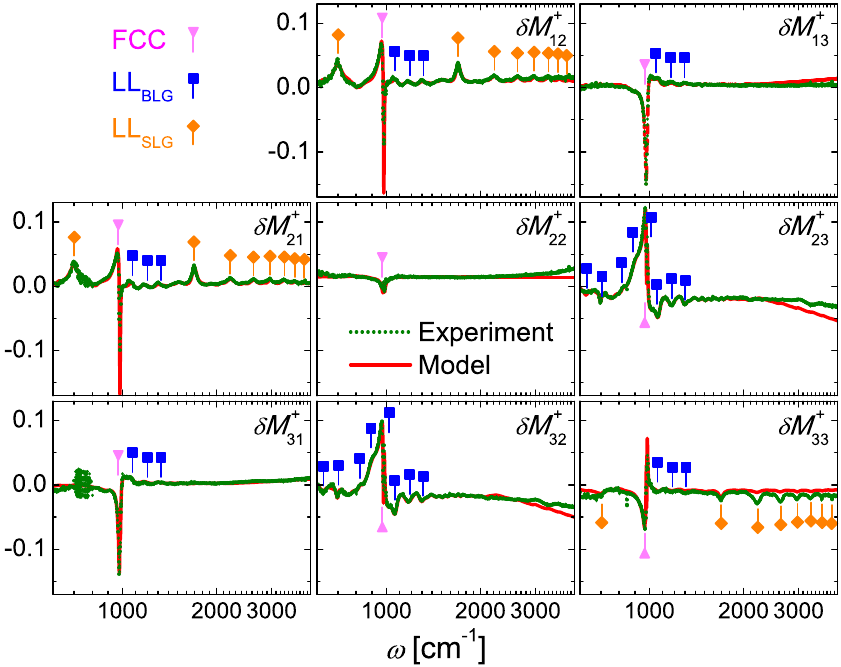}
	\caption{Derived OHE data $\delta\mathbf{M}^{+}$ (green, dotted line) and results from best model calculations (red, solid line) for epitaxial graphene at $\mathit{\Phi}_a$~=~45$^\circ$, $T=1.5$~K and $B=8$~T (effective field parallel to sample normal $B_c=B/\sqrt{2}\approx5.66$~T). The magneto-optical signal near $\nu=1000$~cm$^{-1}$ labeled FCC (triangles, pink) is assigned to free charge carriers. The features labeled LL$_{\stext{BLG}}$ (rectangles, blue) and the set of peaks labeled LL$_{\stext{SLG}}$ (diamonds, orange) are inter-Landau-level transitions in multi-layer and single-layer graphene, respectively. The contributions of LL$_{\stext{SLG}}$ are limited to the on-diagonal-block elements of the Mueller matrix. Therefore, the polarization selection rules for LL$_{\stext{SLG}}$ are polarization conserving - while the processes leading to FCC and LL$_{\stext{BLG}}$ are polarization mixing.~\cite{KuehnePRL111_2013} }
	\label{fig:Graphene1}
\end{figure}
\begin{figure}[!ht]
	\centering
  \includegraphics[
	width=0.46\textwidth]{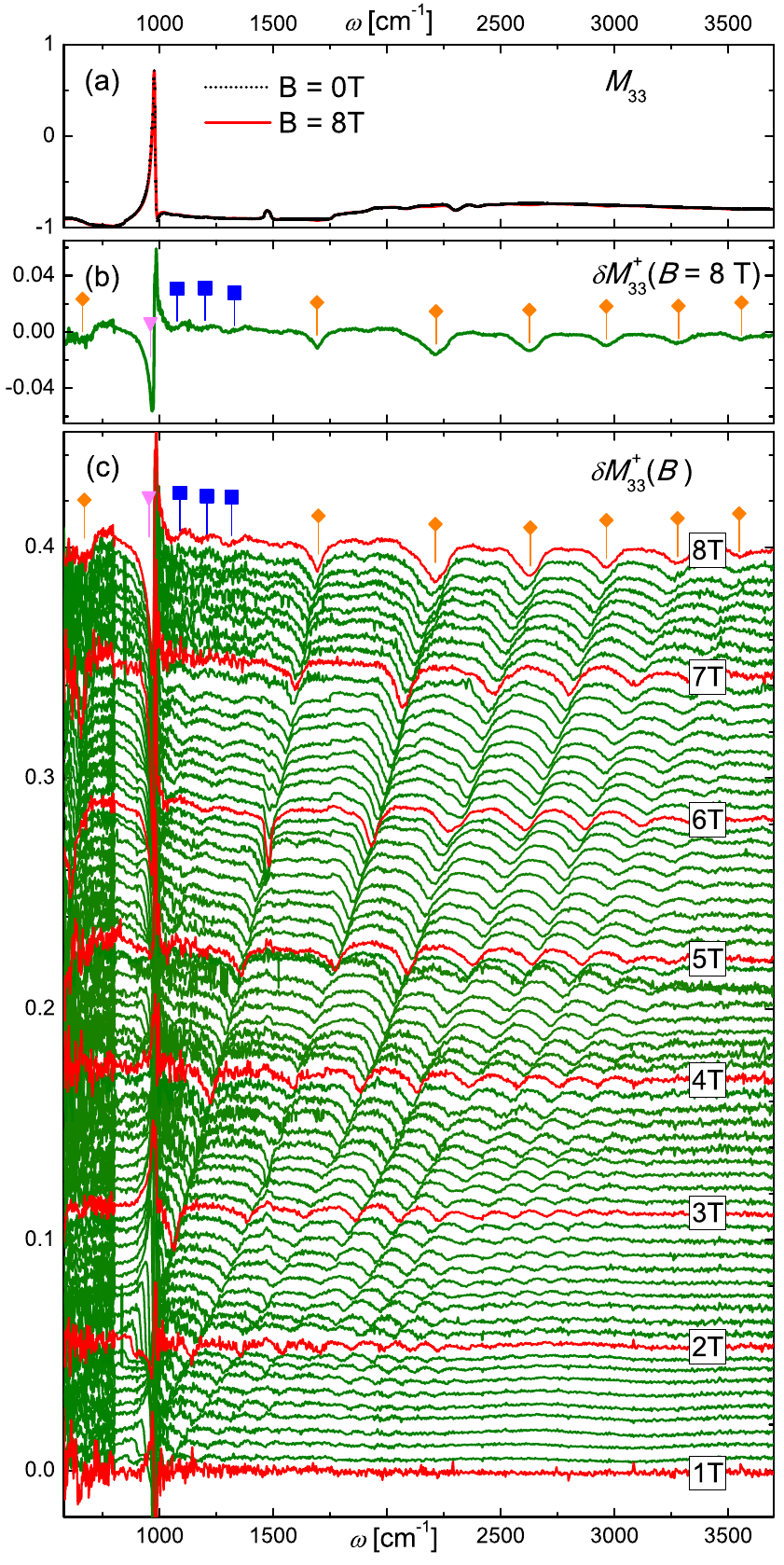}
	\caption{Data for epitaxial graphene at $T=1.5$~K and $\Phi_a$~=~45$^\circ$. (a) Experimental GE data for $B=0$~T (black, dotted line) and OHE data for $B=8$~T (red, solid line), and (b) derived OHE data $\delta\mathbf{M}^{+}$ (difference between the two datasets from (a)) are plotted. Note that the experimental data range for normalized Mueller matrix data is $M_{\text{ij}}=-1\dots 1$, while, here, the OHE only leads to very subtle changes $\left|\delta M_{\text{ij}}\right|\leq 0.05$ with respect to the GE data. (c)~Plot of representative on-diagonal-block Mueller matrix element $\delta M_{33}^+$ of the derived OHE dataset for $B=1\dots 8\text{~T}$, showing SiC substrate phonon mode (near 1000~cm$^{-1}$) enhanced free charge carrier magneto-optic response (triangles, pink), near-linear-$B$ multi-layer graphene inter-Landau-level transitions (rectangles, blue), and the typical $\sqrt{B}$-dependency of inter-Landau-level transitions in single layer graphene (diamonds, orange). Further information can be found in Ref.~\onlinecite{KuehnePRL111_2013}.}
	\label{fig:Graphene2}
\end{figure}
Exemplarily, epitaxial graphene on 6\textit{H}-SiC was investigated to demonstrate the MIR ellipsometer sub-system of the OHE instrument~(Sec.~\ref{sec:MIR}). The epitaxial graphene sample was grown in an argon atmosphere at 1400~$^\circ$C, by sublimating Si from the polar \textit{c}-face (000$\bar{1}$) of a semi-insulating 6\textit{H}-SiC substrate. From previous measurements on C-face 4\textit{H}-SiC~\cite{BoosalisAPL101_2012}, the number of graphene layers is estimated to be 10-20. Further details on growth conditions are beyond the scope of this manuscript, and can be found in Ref.~\onlinecite{TedescoAPL96_2010}.

The OHE experiment was conducted at an angle of incidence of $\Phi_a$~=~45$^\circ$, while the magnetic field was aligned along the reflected beam. Experimental data was recorded in the spectral range from 600 to 4000~cm$^{-1}$ with a spectral resolution of 1~cm$^{-1}$, using the HgCdTe detector, while the sample was kept at a temperature of $T=1.5$~K.

Figure~\ref{fig:Graphene1} shows derived OHE data $\delta\mathbf{M}^{+}$ (green, dotted line) for $B=8$~T (effective field parallel to sample normal $B_c=|\mathbf{B}|/\sqrt{2}\sim 5.66$~T) and results from best model calculations (red, solid line). The model free analysis approach provides valuable information. The OHE data shows several resonances, which can be divided in three groups: a peak near $\nu=1000$~cm$^{-1}$ labeled FCC (triangles, pink), a set of features labeled LL$_{\stext{BLG}}$ (rectangles, blue) and a set of peaks labeled LL$_{\stext{SLG}}$ (diamonds, orange). While the resonances labeled FCC and LL$_{\stext{BLG}}$ are present in all Mueller matrix elements, the features labeled LL$_{\stext{SLG}}$ are only present in the on-diagonal-block elements of the Mueller matrix. Since non-vanishing off-diagonal-block elements in the Mueller matrix are inherently tied to non-vanishing off-diagonal elements in the underlying dielectric tensor, the dielectric tensor has to be diagonal $(\varepsilon_{xy}=\varepsilon_{yx}=0)$. Furthermore, the magneto-optic contributions of LL$_{\stext{SLG}}$ to the permittivity tensor in its representation for circularly polarized light [Eq.~(\ref{eq:LLTeps})] must satisfy $\chi_{\mathrm{+}}=\chi_{\mathrm{-}}$. In other words, the physical processes leading to the fingerprints labeled LL$_{\stext{SLG}}$ are polarization conserving - while the processes leading to FCC and LL$_{\stext{BLG}}$ are polarization mixing (polarization selection rules).

\begin{figure*}[htb]
	\centering
  \includegraphics[
	width=1\textwidth]{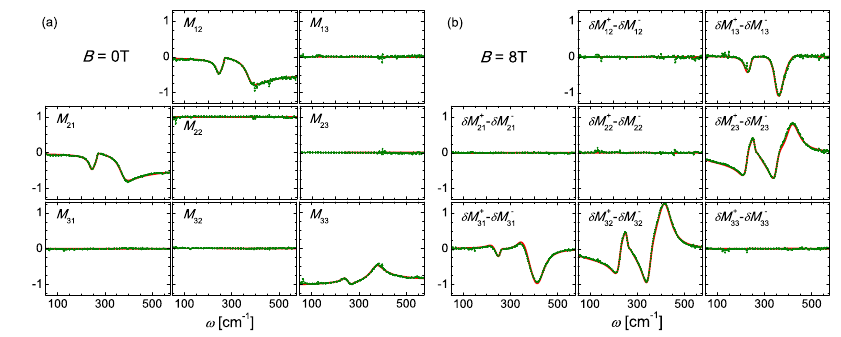}
	\caption{Experimental data (dotted line, green) and best-model calculations (solid line, red) for a GaAs substrate in the FIR spectral range. Non magnetic $(B=0)$~T GE data (a) possesses only non-vanishing on-diagonal-block elements, while in (b) the derived OHE data $\delta\mathbf{M}^{+}-\delta\mathbf{M}^{-}$ (for $B=8$~T) only off-diagonal-block elements are different from zero, indicating the anti-symmetry of the magneto-optic dielectric tensor. The asymmetry between corresponding matrix elements with interchanged indices is caused by the asymmetric orientation of the magnetic field (with respect to the sample normal).}
	\label{fig:GaAs_8T}
\end{figure*}
The origin of the individual processes can be determined by field-dependent measurements. Figure~\ref{fig:Graphene2}~(c) displays the magnetic field dependence of a representative on-diagonal-block Mueller matrix element $\delta M_{33}^{+}$ of the derived OHE dataset for $B=1\dots 8\text{~T}$ in $0.1\text{~T}$ increments (the corresponding representative off-diagonal-block Mueller matrix element is omitted here and the interested reader is referred to Ref.~\onlinecite{KuehnePRL111_2013}). At a first glance it can be noted that the resonance labeled FCC increases in amplitude while its spectral position is not affected. This indicates that the resonance labeled FCC is the SiC phonon mode (best model parameter: $l=1$, \mbox{$\varepsilon_{\infty,{\stext{x,y}}}=6.00$}, \mbox{$\gamma_{\stext{LO},{\stext{x,y}}}=4.74$~cm$^{-1}$}, \mbox{$\omega_{\stext{LO},{\stext{x,y}}}=972.72$~cm$^{-1}$}, \mbox{$\gamma_{\stext{TO},{\stext{x,y}}}=1.18$~cm$^{-1}$}, \mbox{$\omega_{\stext{TO},{\stext{x,y}}}=799.31$~cm$^{-1}$}, \mbox{$\varepsilon_{\infty,{\stext{z}}}=5.84$}, \mbox{$\gamma_{\stext{LO},{\stext{z}}}=2.64$~cm$^{-1}$}, \mbox{$\omega_{\stext{LO},{\stext{z}}}=966.99$~cm$^{-1}$}, \mbox{$\gamma_{\stext{TO},{\stext{z}}}=0.50$~cm$^{-1}$}, and \mbox{$\omega_{\stext{TO},{\stext{z}}}=798.73$~cm$^{-1}$}) enhanced, magneto-optic fingerprint from free charge carriers. In contrast, the spectral positions of the resonances labeled LL$_{\stext{SLG}}$ shift to higher energies and their amplitudes increase with increasing magnetic field strength. A detailed analysis of the peaks reveals a $\sqrt{B}$-dependence of their spectral positions, indicative for inter-Landau-level transitions in the Dirac-type band structure of single-layer graphene with the Fermi level close to the charge neutrality point~\cite{SadowskiPRL97_2006}. Landau level energies in single layered graphene are given by $E_{\stext{SLG}}^{\stext{LL}}(n) =\mbox{sign}(n)E_0\sqrt{|n|}$ with $E_0=\tilde{c}\sqrt{2\hbar e|B_c|}$, where $\tilde{c}$ is the average velocity of Dirac fermions in graphene. Optical selection rules for transitions between levels $n'$ and $n$ demand $|n'|=|n|\pm 1$. The Fermi velocity is determined as $\tilde{c}=(1.01\pm0.01)\times 10^6$m/s, in very good agreement with Refs.~\onlinecite{OrlitaPRB83_2011, SadowskiPRL97_2006, OrlitaPRL102_2009, HenriksenPRL100_2008, OrlitaPRL107_2011, OrlitaPRL101_2008}. The analysis of the magnetic field dependence of the resonances labeled LL$_{\stext{BLG}}$ reveals their origin from the bi-layer branch of inter-Landau-level transitions in ABA-stacked (Bernal) multi-layered graphene. Inter-Landau-level transitions from bi-layer and tri-layer graphene were identified. For further details the interested reader is referred to Ref.~\onlinecite{KuehnePRL111_2013}.

\subsection{The FIR optical Hall effect --- GaAs}\label{sec:GaAs}
The \textit{n}-type GaAs substrate investigated here, was moderately doped with tellurium ($N\approx 10^{18}$~cm$^{-3}$) and is opaque to FIR radiation. All Measurements were conducted at an angle of incidence of $\Phi_a=45^{\circ}$, while the magnetic field was parallel to the reflected beam. Figure~\ref{fig:GaAs_8T} shows experimental data from the FIR spectral range (50--650 cm$^{-1}$) obtained at a sample temperature of $T=300$~K. The sub-figures~(a) and (b) display GE data ($B=0$~T) and derived OHE data $\delta\mathbf{M}^{+}-\delta\mathbf{M}^{-}$ for $B=8$~T, respectively. Without magnetic field, GaAs is optically isotropic, i.e., for $B=0$~T all off-diagonal-block elements ($M_{13}$, $M_{23}$, $M_{31}$, and $M_{32}$) in (a) are zero. In contrast, due to the underlying magneto-optic dielectric tensor symmetry, all on-diagonal-block elements in (b) are zero, indicating the anti-symmetry of the off-diagonal elements of the dielectric tensor, as well as their sign change under field inversion. Note that for a bulk material and a field orientation that is not perpendicular to the sample surface, all off-diagonal elements of the dielectric tensor are different from zero. This also leads to the asymmetry between corresponding Mueller matrix elements with interchanged indices ($ij$ and $ji$).

The optical model for the GaAs substrate consists of a single, semi-infinite layer which contains two contributions, the optical response of polar lattice vibrations and free charge carriers described by the Drude model. Due to the crystal symmetry of GaAs, the dielectric tensor for the lattice vibration [Eq.~(\ref{eqn:phonon2}) and (\ref{Substrate})] has identical diagonal elements, \textit{e.g.} $\varepsilon_{x}^{\text{\tiny{L}}}=\varepsilon_{y}^{\text{\tiny{L}}}=\varepsilon_{z}^{\text{\tiny{L}}}$ and possesses only one TO-LO resonance in the FIR spectral range. Best model parameter for the lattice vibrations were determined as $\varepsilon_{\infty}=(9.269\pm 0.009)$, $\omega_{\text{\tiny{TO}}}=(268.24\pm 0.06)$~cm$^{-1}$, $\omega_{\text{\tiny{LO}}}=(290.96\pm 0.07)$~cm$^{-1}$ and $\gamma_{\text{\tiny{TO}}}=\gamma_{\text{\tiny{LO}}}=(1.9\pm 0.2)$~cm$^{-1}$. The model parameters for the free-charge-carrier contribution $\log N=(17.926\pm 0.001)$~cm$^{-3}$, $\mu=(1789\pm 11)$~cm$^{2}$/Vs and $m=(0.0738\pm 0.0001)$~m$_{\text{e}}$ are in good agreement with results in previous publications.\cite{0022-3719-12-12-014,blakemore:R123}

\subsection{The THz optical Hall effect --- HEMT}\label{sec:HEMT}
The HEMT structure investigated here, was grown on a semi-insulating 4\textit{H}-SiC substrate by metal-organic chemical vapor deposition at temperatures of 1050~$^{\circ}$C.\cite{Forsberg20093007,KakanakovaGeorgieva2007100} First, a nominally 100~nm thick AlN nucleation layer was grown, followed by a 1800~nm thick GaN buffer and a 20~nm thick Al$_{0.25}$Ga$_{0.75}$N electron barrier layer.

\begin{figure}[!t]
	\centering
  \includegraphics[
	width=0.48\textwidth]{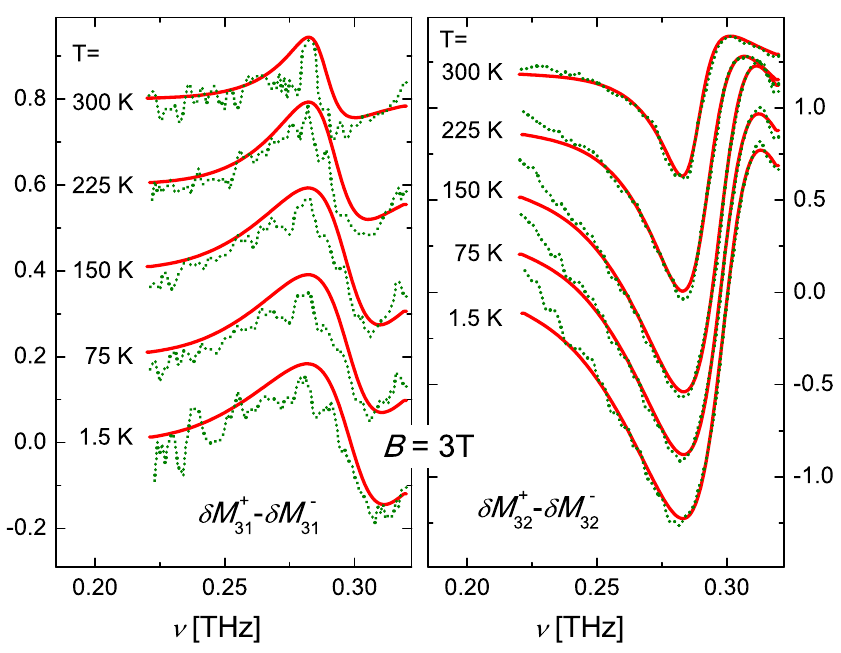}
	\caption{AlGaN/GaN HEMT structure: experimental (dotted lines, green) and best-model calculated (solid lines, red) OHE data for $B=3$~T ($\delta{M}_{31}^{+}-\delta{M}_{31}^{-}$, left panel; $\delta{M}_{32}^{+}-\delta{M}_{32}^{-}$, right panel) between 0.22 and 0.32~THz. The spectra were obtained for temperatures ranging from 1.5 to 300~K, at an angle of incidence $\Phi_a=45^{\circ}$, and show the Fabry-P\'erot interference enhanced magneto-optic response of free charge carriers. The spectra for $T=75$~K and above are stacked by 0.1.}
	\label{fig:HEMT}
\end{figure}

Figure~\ref{fig:HEMT} displays the temperature dependence of non-vanishing off-diagonal-block Mueller matrix elements of derived OHE data $\delta M_{31}^{+}-\delta M_{31}^{-}$ and $\delta M_{32}^{+}-\delta M_{32}^{-}$ for $B=3$~T in the spectral range from 0.22 to 0.32~THz. The experimental OHE data is depicted as dotted lines~(green), while best-model calculated OHE data is plotted as solid lines~(red) for different temperatures between 1.5~K and 300~K. The observed OHE is caused by free charge carriers in the high mobility channel of the HEMT structure, which is enhanced by the Fabry-P\'erot interference in the $(356\pm 1)$~$\mu$m thick SiC substrate. The optical model is composed of a SiC substrate layer, a AlN nucleation layer, a GaN buffer layer, a layer for the GaN HEMT channel and a AlGaN layer. The thickness of the GaN HEMT layer was set to $d=1$~nm and not varied during data analysis. The HEMT layer thickness was used to calculate the sheet charge density $N_s=N d$, where $N$ is the bulk charge density. Derived THz OHE data was analyzed simultaneously with GE data recorded with a commercial MIR ellipsometer.\cite{HofmannAPL_2012} During model based OHE data analysis the parameters for the sheet charge density $N_s$, the mobility $\mu$ and the effective mass parameter $m$ were determined. The sheet carrier density was found to be constant within the error limits at a value of $N_s=N d=(5\pm 1) \times 10^{12}$~cm$^{-2}$ for all temperatures. The mobility and the effective mass parameters were determined as $m=(0.22\pm 0.01) m_0$ ($m=(0.36\pm 0.03) m_0$) and $\mu=(7800\pm 410)$~cm$^{2}$/Vs ($\mu=(1711\pm 150)$~cm$^{2}$/Vs) at $T=1.5$~K ($T=300$~K), where $m_0$ is the mass of the free electron. Further details are omitted here and the interested reader is referred to an in depth discussion in Ref.~\onlinecite{HofmannAPL_2012}.

\section{Summary}\label{sec:Summary}
In this article, we have given an overview over theoretical and experimental aspects of the OHE, and have successfully demonstrated the operation of an integrated MIR, FIR and THz optical Hall effect instrument. 
Two in-house built ellipsometers sub-systems, operating in the rotating-analyzer configuration, were employed to determine the upper $3\times 3$ block of the normalized Mueller matrix in an ultra wide spectral range from 3~cm$^{-1}$ to 7000~cm$^{-1}$ (0.1--210~THz or 0.4--870~meV). Different aspects of the integrated MIR, FIR and THz optical Hall effect instrument, such as the cryogen-free, superconducting 8~Tesla magneto-cryostat sub-system, and the optical setup of the integrated MIR, FIR and THz optical Hall effect instrument, as well as the data acquisition and data analysis strategies were discussed in detail. For demonstration purposes of the operation of the integrated MIR, FIR and THz optical Hall effect instrument in the MIR, FIR and THz spectral range, three sample systems were studied: an epitaxial graphene sample (MIR), a GaAs substrate (FIR) and a AlGaN/GaN HEMT structure (THz). The presented data covered the full magnetic field range ($B=0\dots 8$~T) and temperature range ($T=1.5\dots 300$~K) of the system. We have demonstrated that the integrated MIR, FIR and THz optical Hall effect instrument can be used to determine the parameters of free charge carriers in bulk, and in two-dimensional confinement. Furthermore, it was shown that the integrated MIR, FIR and THz optical Hall effect instrument is capable to investigate the symmetry properties of magneto-optic, dielectric tensors of quantized systems, and was able to determine the polarization selection rules of inter-Landau-level transitions. The integrated MIR, FIR and THz optical Hall effect instrument has a wide range of applications, from determining free charge carrier properties contact-free, over the access to free charge carrier properties in buried layers, to magneto-optic quantum phenomena, polarization selection rules, to symmetry properties of magneto-optic dielectric tensors.

\section{Acknowledgments}\label{sec:Acknowledgments}
This work was supported by the Army Research Office (D. Woolard, Contract No. W911NF-09-C-0097), the National Science Foundation (Grant Nos. MRSEC DMR-0820521, DMR-0907475, EPS-1004094), with primary support under MRI DMR-0922937, the University of Nebraska-Lincoln, the J.A.~Woollam Foundation and the J.A.~Wollam Company. The authors thank Dr.~D.K.~Gaskill and his workgroup at the U.S.~Naval Research Laboratory, Washington DC for providing epitaxial graphene samples. The authors would like to acknowledge Dr.~V.~Darakchieva and Dr.~E.~Jansen and their workgroups at the University of Link\"{o}ping, Sweden for providing the HEMT structure samples. Thanks to C.~Rice, C.~Briley, S.~M.~Slone and A.~Boosalis for their help building and testing mechanical and electrical components. Special thanks to the instruments shop at the department of physics at UNL, and in particular to L.~Marquart for his invaluable advice.



\begin{thebibliography}{102}
\bibitem [{RSI({\natexlab{a}})}]{RSI_Purge_comment}%
  \BibitemOpen
  \bibinfo {note} {With exception of the
  space between the polarization preparation unit, the sample and the
  polarization detection unit, which is either filled with not-dryed air, or,
  if the magneto-cryostat sub-system is used, filled with ultra high purity
  helium gas or is under vacuum.}\BibitemShut {Stop}%
\end{thebibliography}
%


\end{document}